\documentclass[sigconf,nonacm]{acmart}

\usepackage{subfigure}
\usepackage{siunitx}
\usepackage{enumitem}

\usepackage{algorithm}
\usepackage[noend]{algpseudocode}
\let\ReturnInline\Return
\renewcommand{\Return}{\State\ReturnInline}
\algrenewcommand\algorithmicrequire{$\rhd$}
\algrenewcommand\algorithmicensure{$\square$}

\AtBeginDocument{%
  \providecommand\BibTeX{{%
    \normalfont B\kern-0.5em{\scshape i\kern-0.25em b}\kern-0.8em\TeX}}}

\setcopyright{acmcopyright}
\copyrightyear{2018}
\acmYear{2018}
\acmDOI{XXXXXXX.XXXXXXX}

\acmConference[Conference acronym 'XX]{Make sure to enter the correct
  conference title from your rights confirmation emai}{June 03--05,
  2018}{Woodstock, NY}

\newcommand{\ignore}[1]{}

\begin{document}

\title[GSL-LPA: Fast Label Propagation Algorithm (LPA) for Community Detection with no Internally-Disconnected Communities]{GSL-LPA: Fast Label Propagation Algorithm (LPA) for Community Detection with no Internally-Disconnected Communities}


\author{Subhajit Sahu}
\email{subhajit.sahu@research.iiit.ac.in}
\affiliation{%
  \institution{IIIT Hyderabad}
  \streetaddress{Professor CR Rao Rd, Gachibowli}
  \city{Hyderabad}
  \state{Telangana}
  \country{India}
  \postcode{500032}
}


\settopmatter{printfolios=true}

\begin{abstract}
Community detection is the problem of identifying tightly connected clusters of nodes within a network. Efficient parallel algorithms for this play a crucial role in various applications, especially as datasets expand to significant sizes. The Label Propagation Algorithm (LPA) is commonly employed for this purpose due to its ease of parallelization, rapid execution, and scalability --- however, it may yield internally disconnected communities. This technical report introduces GSL-LPA, derived from our parallelization of LPA, namely GVE-LPA. Our experiments on a system\ignore{equipped} with two 16-core Intel Xeon Gold 6226R processors show that GSL-LPA not only mitigates this issue but also surpasses FLPA, igraph LPA, and NetworKit LPA by $55\times$, $10,300\times$, and $5.8\times$, respectively, achieving a processing rate of $844M$ edges/s on a $3.8B$ edge graph. Additionally, GSL-LPA scales at a rate of $1.6\times$ for every doubling of threads.
\end{abstract}



\maketitle

\section{Introduction}
\label{sec:introduction}
Community detection refers to the task of identifying groups of vertices characterized by dense internal connections and sparse connections between groups \cite{com-fortunato10}. These groups, often referred to as communities or clusters \cite{abbe2018community}, offer valuable insights into the structure and function of networks \cite{com-fortunato10}. Community detection has a wide range of applications across multiple fields. In ecology, it is applied to assess if food webs are structured into compartments, where species within each compartment interact more frequently with each other than with those in different compartments \cite{guimera2010origin}. Within healthcare, it helps in analyzing the dynamics of populations vulnerable to epidemic diseases \cite{salathe2010dynamics}, detecting conditions such as lung cancer \cite{bechtel2005lung}, and classifying tumor types using genomic data \cite{haq2016community}. In drug discovery, community detection aids in identifying groups of similar compounds or target proteins, which supports the discovery of novel therapeutic agents \cite{ma2019comparative}. Furthermore, community detection has been effectively used to reveal the structure and evolution of complex biological networks, including metabolic networks \cite{kim2009centralized}, Gene Regulatory Networks (GRNs) \cite{rivera2010nemo}, and Lateral Gene Transfer (LGT) networks \cite{popa2011directed}. It also plays a significant role in understanding human brain networks, enabling insights into their structural and functional organization \cite{he2010graph}.

The problem of community detection is due to the lack of prior knowledge about the number of communities and their size distribution \cite{com-gregory10}. Heuristic-based approaches are commonly employed for community detection \cite{com-raghavan07, com-blondel08, com-xie11, traag2023large, clauset2004finding, duch2005community, reichardt2006statistical, com-kloster14, com-traag19, com-you20, com-rosvall08, com-whang13}. The modularity metric \cite{com-newman06} is used to measure the quality of communities identified. The communities identified are considered intrinsic when based solely on network topology and disjoint when each vertex belongs to only one community \cite{com-gregory10, coscia2011classification}. The Label Propagation Algorithm (LPA), also known as RAK \cite{com-raghavan07}, is a widely adopted diffusion-based heuristic for community detection, capable of identifying communities of moderate quality. It offers simplicity, speed, and scalability advantages over the Louvain method \cite{com-blondel08}, another effective community detection algorithm renowned for its high-quality results. Specifically, we observe LPA to outpace Louvain by $2.3 - 14\times$ in terms of speed, while identifying communities of $3 - 30\%$ lower quality. LPA's efficiency is attributed to its avoidance of repeated optimization steps and its ease of parallelization. Thus, LPA is well-suited for applications prioritizing high performance at the expense of slightly lower result quality. While LPA typically yields communities with lower modularity scores, it has been observed to achieve high Normalized Mutual Information (NMI) score relative to the ground truth \cite{peng2014accelerating}. In our experimentation with alternative label-propagation based methods such as COPRA \cite{com-gregory10}, SLPA \cite{com-xie11}, and LabelRank \cite{com-xie13}, we found LPA to be the most efficient, while identifying communities of equivalent quality \cite{sahu2023selecting}.

However, LPA is susceptible to identifying a large number of internally disconnected communities. Through our experimental analysis, we observe that up to $3.1\%$, $0.3\%$, and $14.5\%$ of communities identified using FLPA \cite{traag2023large}, igraph LPA \cite{csardi2006igraph}, and NetworKit LPA \cite{staudt2016networkit} exhibit this issue. Rectifying such disconnected communities is crucial to ensuring the accuracy and robustness of community detection algorithms. While a number of studies have proposed addressing internally-disconnected communities as a post-processing step, they do not present a concrete algorithm to do so \cite{com-raghavan07, com-gregory10, hafez2014bnem, luecken2016application, wolf2019paga}. In this technical report, we present a GSL-LPA, which is based upon an efficient multicore implementation of LPA \cite{sahu2023gvelpa}, and utilizes our BFS-based approach for splitting internally-disconnected communities --- which we show, outperforms other approaches for splitting the identified disconnected communities.

\section{Related work}
\label{sec:related}
\ignore{Communities within networks serve as functional units or meta-nodes \cite{com-chatterjee19}. Identifying these divisions in an unsupervised manner is crucial in various domains, including drug discovery, disease prediction, protein annotation, topic discovery, link prediction, recommendation systems, customer segmentation, inferring land use, and criminal identification \cite{com-karatas18}.}A number of schemes have been devised for community detection \cite{com-raghavan07, com-blondel08, com-xie11, traag2023large, clauset2004finding, duch2005community, reichardt2006statistical, com-rosvall08, com-whang13, com-kloster14, com-traag19, com-you20}. These can be categorized into three basic approaches: bottom-up, top-down, and data-structure based \cite{com-souravlas21}. Additionally, they can be classified as divisive, agglomerative, or multi-level methods \cite{com-zarayeneh21}. To evaluate these methods, fitness scores like modularity \cite{newman2006modularity} are used.\ignore{Modularity measures the relative density of links within communities compared to those outside and ranges from $-0.5$ to $1$.}\ignore{Optimizing modularity theoretically leads to the best possible clustering of nodes \cite{com-lancichinetti09}. However, exhaustive exploration of all possible node clusterings is impractical \cite{com-fortunato10}, prompting the use of heuristic methods like the Label Propagation Algorithm (LPA) \cite{com-raghavan07}.}

The Label Propagation Algorithm (LPA) is a diffusion-based method for identifying communities. Techniques for parallelizing LPA include vertex assignment with guided scheduling \cite{staudt2015engineering}, parallel bitonic sort \cite{soman2011fast}, and pre-partitioning of the graph \cite{kuzmin2015parallelizing}.\ignore{LPA considers only adjacent nodes for label selection. Alternative neighborhood definitions for LPA consider nodes at a distance less than or equal to 2 \cite{lou2013detecting, chen2017detecting} or use a tunable parameter for distance \cite{sun2014label}.} In terms of quality, LPA may yield communities with low modularity score, resulting in the emergence of a dominant community that envelops the majority of the nodes, obscuring finer community structures \cite{com-gregory10}. To address instability of result, community structure quality, and performance due to vertex processing order, a few ordering strategies have been attempted. These include updating only the active nodes \cite{xie2011community} or prioritizing specific subsets based on properties, like core or boundary nodes \cite{gui2018community}. In some cases, oscillations in label assignments may arise instead of convergence. To address this, asynchronous mode \cite{leung2009towards}, alternating updates of independent node subsets \cite{cordasco2012label}, and parallel graph coloring techniques \cite{cordasco2012label} have been attempted. Raghavan et al. \cite{com-raghavan07} note that after just five iterations, labels of $95\%$ of nodes converge to their final values.

Additional enhancements to LPA encompass employing a stable (non-random) mechanism for selecting labels when encountering multiple best labels \cite{com-xing14}, mitigating the challenge posed by monster communities \cite{com-berahmand18, com-sattari18}, discerning central nodes and combining communities to enhance modularity \cite{com-you20}, as well as leveraging frontiers with alternating push-pull to minimize the number of edges traversed and enhance solution quality \cite{com-liu20}. While a number of variations of LPA have been introduced, the original formulation remains the simplest and most efficient \cite{garza2019community}.

A few open-source implementations and software packages have been developed for community detection utilizing the LPA. Fast Label Propagation Algorithm (FLPA) \cite{traag2023large} is a variant of the LPA that employs a queue-based method, processing only vertices with recently updated neighborhoods. NetworKit \cite{staudt2016networkit} is a software package built to analyze the structural attributes of graph datasets containing billions of connections. It integrates C++ kernels with a Python frontend, and features a parallel implementation of LPA. igraph \cite{csardi2006igraph} is a similar package that is written in C, and offers Python, R, and Mathematica frontends. It has good adoption, and includes an LPA implementation.

Raghavan et al. \cite{com-raghavan07} note that their Label Propagation Algorithm (LPA) for community detection can uncover internally-disconnected communities \cite{com-gregory10}. Genetic algorithms \cite{hesamipour2022detecting} and expectation minimization/maximization algorithms \cite{ball2011efficient, hafez2014bnem} can also detect such disconnected communities. To address this, they propose applying Breadth-First Search (BFS) to subnetworks in each individual group to separate the disconnected communities. This can be achieved with a time complexity of $O(M+N)$. Gregory \cite{com-gregory10} extends LPA with the Community Overlap PRopagation Algorithm (COPRA), which removes nested communities and splits disconnected communities using a method similar to Raghavan et al. \cite{com-raghavan07}. Hafez et al. \cite{hafez2014bnem} introduce a community detection algorithm using Bayesian Network and Expectation Minimization (BNEM), which includes a step to detect disconnected components within communities. If detected, disconnected components are assigned new community labels. This scenario arises when the network has more communities than the specified number $k$ in the algorithm. Hesamipour et al. \cite{hesamipour2022detecting} propose a genetic algorithm for community detection, integrating similarity-based and modularity-based approaches. They use an MST-based representation to address issues like disconnected communities and ineffective mutations.\ignore{Goel et al. \cite{goel2023effective} present a tangential work, where they identify a set of nodes in a network, known as a Structural Hole Spanner (SHS), that act as a bridge among different otherwise disconnected communities.}

\section{Preliminaries}
\label{sec:preliminaries}
Consider an undirected graph $G(V, E, w)$, where $V$ denotes the set of vertices, $E$ denotes the set of edges, and $w_{ij} = w_{ji}$ denotes the weight associated with each edge ($i$ and $j$ represent the vertices at the endpoints of each edge). In the case of unweighted graphs, we assume a unit weight for each edge, i.e., $w_{ij} = 1$. The neighbors of a vertex $i$ are represented as $J_i = \{j\ |\ (i, j) \in E\}$, the weighted degree of each vertex is represented as $K_i = \sum_{j \in J_i} w_{ij}$, the total number of vertices is $N = |V|$, the total number of edges is $M = |E|$, and the sum of edge weights in the undirected graph is $m = \sum_{i, j \in V} w_{ij}/2$.

\subsection{Community detection}

Disjoint community detection involves assigning a community ID $c$ from set $\Gamma$ to each vertex $i \in V$, denoted by the mapping $C: V \rightarrow \Gamma$. Vertices belonging to community $c \in \Gamma$ are denoted as $V_c$, and the community of a vertex $i$ is represented by $C_i$. Further, we represent the neighbors of vertex $i$ within community $c$ as $J_{i \rightarrow c} = \{j\ |\ j \in J_i\ \text{and}\ C_j = c\}$, their corresponding edge weights summed as $K_{i \rightarrow c} = \sum_{j \in J_{i \rightarrow c}} w_{ij}$, the edge weight within community $c$ as $\sigma_c = \sum_{(i, j) \in E\ \text{and}\ C_i = C_j = c} w_{ij}$, and the total edge weight of community $c$ as $\Sigma_c = \sum_{(i, j) \in E\ \text{and}\ C_i = c} w_{ij}$ \cite{com-leskovec21}.

\subsection{Modularity}

Modularity is a metric that is used for assessing the quality of communities detected by heuristic-based algorithms. It quantifies the difference between the observed fraction of edges within communities and the expected fraction under random edge distribution. Modularity values range from $-0.5$ to $1$, where higher values indicate better community structure \cite{com-brandes07}.\ignore{The optimization of this metric theoretically leads to the optimal grouping \cite{com-newman04, com-traag11}.} The modularity $Q$ is calculated using Equation \ref{eq:modularity}, where $\delta$ represents the Kronecker delta function ($\delta (x,y)=1$ if $x=y$, $0$ otherwise$)$.\ignore{The \textit{delta modularity} of moving a vertex $i$ from community $d$ to community $c$, denoted as $\Delta Q_{i: d \rightarrow c}$, can be computed using Equation \ref{eq:delta-modularity}.}

\begin{equation}
\label{eq:modularity}
  Q
  = \frac{1}{2m} \sum_{(i, j) \in E} \left[w_{ij} - \frac{K_i K_j}{2m}\right] \delta(C_i, C_j)
  = \sum_{c \in \Gamma} \left[\frac{\sigma_c}{2m} - \left(\frac{\Sigma_c}{2m}\right)^2\right]
\end{equation}

\ignore{\begin{equation}
\label{eq:delta-modularity}
  \Delta Q_{i: d \rightarrow c}
  = \frac{1}{m} (K_{i \rightarrow c} - K_{i \rightarrow d}) - \frac{K_i}{2m^2} (K_i + \Sigma_c - \Sigma_d)
\end{equation}}

\subsection{Label Progagation Algorithm (LPA)}
\label{sec:about-rak}

LPA \cite{com-raghavan07} is a widely used diffusion-based technique for identifying medium-quality communities in large networks. It is known for its simplicity, speed, and scalability\ignore{compared to the Louvain method \cite{com-blondel08}}. In LPA, each vertex $i$ initially possesses a unique label (community ID) $C_i$. During each iteration, vertices adopt the label with the highest interconnecting weight, as illustrated in Equation \ref{eq:lpa}. This iterative process leads to a consensus among densely connected groups of vertices. The algorithm converges when at least $1-\tau$ fraction of vertices (where $\tau$ is the iteration tolerance parameter) maintain their community membership. LPA exhibits a time complexity of $O(L |E|)$ and a space complexity of $O(|V| + |E|)$, with $L$ representing the number of iterations performed \cite{com-raghavan07}. LPA has also been utilized in influence maximization, where the goal is to identify a small set of influential nodes in a network that can maximize the spread of information \cite{kumar2021elpr}.

\begin{equation}
\label{eq:lpa}
  C_i =\ \underset{c\ \in \ \Gamma}{\arg\max} { \sum_{j \in J_i\ |\ C_j = c} w_{ij} }
\end{equation}

\begin{figure*}[hbtp]
  \centering
  \subfigure[The initial community structure with $4$ communities]{
    \label{fig:onrak--1}
    \includegraphics[width=0.31\linewidth]{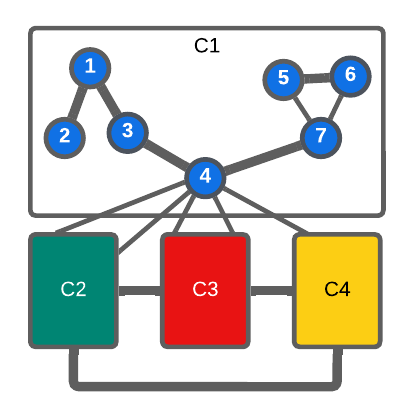}
  }
  \subfigure[After a few iterations,\ignore{communities} $C2$, $C3$, and $C4$ combine]{
    \label{fig:onrak--2}
    \includegraphics[width=0.31\linewidth]{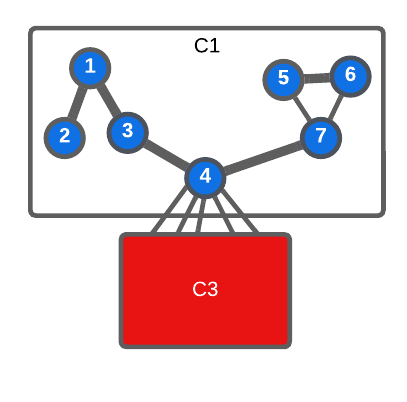}
  }
  \subfigure[Following this, vertex $4$ transitions to $C3$]{
    \label{fig:onrak--3}
    \includegraphics[width=0.31\linewidth]{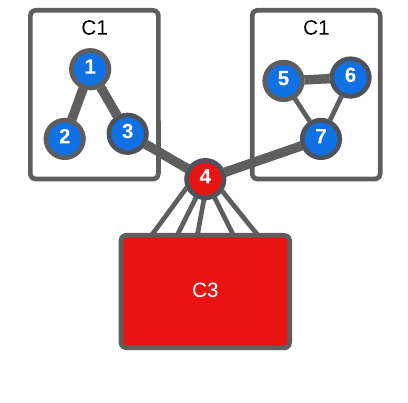}
  } \\[-2ex]
  \caption{An example illustrating the potential of internally-disconnected communities with LPA \cite{sahu2024addressing}. Here, $C1$, $C2$, $C3$, and $C4$ represent four communities derived after one iteration of LPA, with vertices $1$ to $7$ belonging to community $C1$. Thicker lines are used here to indicate higher edge weights.}
  \label{fig:onrak}
\end{figure*}

\subsection{Possibility of Internally-disconnected communities with LPA}
\label{sec:about-rak-disconnected}

LPA has been noted to potentially identify internally disconnected communities \cite{com-raghavan07}. This is illustrated with an example in Figure \ref{fig:onrak}. Initially, Figure \ref{fig:onrak--1} show the community structure after one iteration of LPA, with four communities labeled $C1$, $C2$, $C3$, $C4$, and vertices $1$ to $7$ grouped in $C1$ (i.e., vertices $1$ to $7$ have identified $C1$ as the most weighted label among their neighbors). Subsequent iterations, illustrated in Figure \ref{fig:onrak--2}, result in relabeling of vertices in $C2$ and $C4$ into $C3$, due to strong internal connections, resulting in the combining of communities $C2$, $C3$, and $C4$ into $C3$. This results in $C3$ becoming the most weighted label for vertex $4$. Thus, $4$ changes its label to $C3$, resulting in internal disconnection of $C1$ as vertices $1$, $2$, $3$, $5$, $6$, and $7$ retain their locally popular labels. This phenomenon is common across various LPA implementations, including FLPA \cite{traag2023large}, igraph LPA \cite{csardi2006igraph}, and NetworKit LPA \cite{staudt2016networkit}, which identify up to $3.1\%$, $0.3\%$, and $14.5\%$ of communities that are internally-disconnected, respectively.

\section{Approach}
\label{sec:approach}
In the preceding section, we discussed the occurrence of internally-disconnected communities with LPA. However, this phenomenon is not exclusive to LPA and has been observed in various other community detection algorithms \cite{com-blondel08, hesamipour2022detecting, ball2011efficient, hafez2014bnem}. To address this issue, a commonly used approach involves splitting disconnected communities as a post-processing step (i.e., after the completion of all iterations of the community detection algorithm, when the community memberships of vertices have converged), known as Split Last (SL) using Breadth First Search (BFS) \cite{com-raghavan07, com-gregory10, hafez2014bnem, luecken2016application, wolf2019paga}.

To split the internally-disconnected communities using the SL method, we investigate three different techniques: minimum-label-based Label Propagation (LP), minimum-label-based Label Propagation with Pruning (LPP), and Breadth First Search (BFS). We choose to explore LP and LPP techniques for splitting disconnected communities due to their inherent parallelizability.

Using the LP technique, each vertex in the graph is initially assigned a unique label corresponding to its vertex ID. Subsequently, in every iteration, each vertex within its assigned community selects the minimum label among its neighbors. This iterative process continues until convergence is achieved for all vertex labels. Consequently, each vertex acquires a unique label within its connected component and its respective community, thereby partitioning communities consisting of multiple connected components. On the other hand, the LPP technique integrates a pruning optimization step where only unprocessed vertices are processed. Once a vertex is processed, it is flagged as such and is only reactivated (or marked as unprocessed) if one of its neighbors alter their labels. The pseudocode for both the LP and LPP techniques is provided in Algorithm \ref{alg:splitlp}, with its explanation given in Section \ref{sec:splitlp}.

In contrast, the BFS technique for splitting internally disconnected communities involves selecting a random vertex within each community and identifying all reachable vertices from it as part of one subcommunity. If any vertices remain unvisited in the original community, another random vertex is selected from the remaining set, and the process repeats until all vertices within each community are visited. Consequently, the BFS technique facilitates the partitioning of connected components within each community. The pseudocode for this technique is presented in Algorithm \ref{alg:splitbfs}, with detailed explanations provided in Section \ref{sec:splitbfs}.

\subsection{Explanation of LP/LPP technique\ignore{ for splitting disconnected communities}}
\label{sec:splitlp}

We now present the pseudocode for the parallel minimum-label-based Label Propagation (LP) and Label Propagation with Pruning (LPP) techniques, outlined in Algorithm \ref{alg:splitlp}, which are utilized for partitioning internally-disconnected communities. These techniques serve as a post-processing step (SL) at the end of the community detection algorithm. The function \texttt{splitDisconnectedLp()} accepts as input the graph $G(V, E)$ and the community memberships $C$ of vertices, and outputs the updated community memberships $C'$ where all disconnected communities have been separated.

The algorithm commences in lines \ref{alg:splitlp--init-begin}-\ref{alg:splitlp--init-end} by initializing the minimum labels $C'$ of each vertex to their respective vertex IDs and marking all vertices as unprocessed. Lines \ref{alg:splitlp--loop-begin}-\ref{alg:splitlp--loop-end} denote the iteration loop of the algorithm. Initially, the number of changed labels $\Delta N$ is set (line \ref{alg:splitlp--loopinit}). Subsequently, an iteration of label propagation is performed (lines \ref{alg:splitlp--vertexloop-begin}-\ref{alg:splitlp--vertexloop-end}), followed by a convergence check (line \ref{alg:splitlp--checkconverged}). During each iteration, unprocessed vertices are handled concurrently. For each unprocessed vertex $i$, if the Label Propagation with Pruning (LPP) technique is used, it is marked as processed. The algorithm then identifies the minimum label $c'_{min}$ within the community of vertex $i$ (lines \ref{alg:splitlp--findmin-begin}-\ref{alg:splitlp--findmin-end}) by considering only the labels of its neighbors belonging to the same community, and its own label. If $c'_{min}$ differs from the current minimum label $C'[i]$ (line \ref{alg:splitlp--skipmin}), $C'[i]$ is updated, $\Delta N$ is incremented to mark the change in label, and neighboring vertices of $i$ in the same community are marked as unprocessed (to be updated in the subsequent iteration). The label propagation loop (lines \ref{alg:splitlp--loop-begin}-\ref{alg:splitlp--loop-end}) continues until no further changes occur (line \ref{alg:splitlp--checkconverged}). Finally, the updated labels $C'$, reflecting the community membership of each vertex with no internally-disconnected communities, are returned (line \ref{alg:splitlp--return}).

\begin{algorithm}[hbtp]
\caption{Split disconnected communities using (min) LP \cite{sahu2024addressing}.}
\label{alg:splitlp}
\begin{algorithmic}[1]
\Require{$G(V, E)$: Input graph}
\Require{$C$: Initial community membership/label of each vertex}
\Ensure{$C'$: Updated community membership/label of each vertex}
\Ensure{$c'_{min}$: Minimum connected label within the community}
\Ensure{$\Delta N$: Number of changes in labels}

\Statex

\Function{splitDisconnectedLp}{$G, C$} \label{alg:splitlp--begin}
  \State $C' \gets \{\}$ \label{alg:splitlp--init-begin}
  \ForAll{$i \in V$ \textbf{in parallel}}
    \State Mark $i$ as unprocessed
    \State $C'[i] = i$
  \EndFor \label{alg:splitlp--init-end}
  \Loop \label{alg:splitlp--loop-begin}
    \State $\Delta N \gets 0$ \label{alg:splitlp--loopinit}
    \ForAll{unprocessed $i \in V$ \textbf{in parallel}} \label{alg:splitlp--vertexloop-begin}
      \If{\textbf{is SL-LPP or SP-LPP}}
        \State Mark $i$ as processed
      \EndIf
      \State $\rhd$ Find minimum community label
      \State $c'_{min} \gets C'[i]$ \label{alg:splitlp--findmin-begin}
      \ForAll{$j \in G.out(i)$}
        \If{$C[j] = C[i]$}
          \State $c'_{min} \gets min(C'[j], c'_{min})$
        \EndIf
      \EndFor \label{alg:splitlp--findmin-end}
      \If{$c'_{min} = C'[i]$} \textbf{continue} \label{alg:splitlp--skipmin}
      \EndIf
      \State $\rhd$ Update community label
      \State $C'[i] \gets c'_{min}$ \textbf{;} $\Delta N \gets \Delta N + 1$ \label{alg:splitlp--updatemin--begin}
      \If{\textbf{is SL-LPP or SP-LPP}}
        \ForAll{$j \in G.out(i)$}
          \State Mark $j$ as unprocessed \textbf{if} $C[j] = C[i]$
        \EndFor
      \EndIf \label{alg:splitlp--updatemin--end}
    \EndFor \label{alg:splitlp--vertexloop-end}
    \State $\rhd$ Converged?
    \If{$\Delta N = 0$} \textbf{break} \label{alg:splitlp--checkconverged}
    \EndIf
  \EndLoop \label{alg:splitlp--loop-end}
  \Return{$C'$} \label{alg:splitlp--return}
\EndFunction \label{alg:splitlp--end}
\end{algorithmic}
\end{algorithm}

\subsection{Explanation of BFS technique\ignore{ for splitting disconnected communities}}
\label{sec:splitbfs}

Following that, we move on to describe the pseudocode of the parallel Breadth First Search (BFS) technique, as outlined in Algorithm \ref{alg:splitbfs}, designed for the partitioning of disconnected communities. Similar to the LP/LPP techniques, this approach can be utilized as a post-processing step (SL) at the conclusion of the process. The function \texttt{splitDisconnectedBfs()} receives the input graph $G(V, E)$ and the community membership $C$ of each vertex, and yields the updated community membership $C'$ of each vertex where all the disconnected communities have been separated.

Initially, between lines \ref{alg:splitbfs--init-begin} and \ref{alg:splitbfs--init-end}, the flag vector $vis$, indicating visited vertices, is initialized, and the labels $C'$ for each vertex are set to their respective vertex IDs. Subsequently, multiple thread concurrently process each vertex $i$ in the graph $G$ (lines \ref{alg:splitbfs--loop-begin} to \ref{alg:splitbfs--loop-end}). If the community $c$ of vertex $i$ is not present in the work-list $work_t$ of the current thread $t$, or if vertex $i$ has already been visited, the thread proceeds to the next iteration (line \ref{alg:splitbfs--work}). Conversely, if community $c$ is in the work-list $work_t$ of the current thread $t$ and vertex $i$ has not been visited, a Breadth-First Search (BFS) is executed from vertex $i$ to explore vertices within the same community. This BFS employs lambda functions $f_{if}$ to selectively perform BFS on vertex $j$ if it belongs to the same community, and $f_{do}$ to update the label of visited vertices after each vertex is explored during BFS (line \ref{alg:splitbfs--bfs}). After processing all vertices, the threads synchronize, and the updated labels $C'$, denoting the community membership of each vertex with no disconnected communities, are returned (line \ref{alg:splitbfs--return}). It is worth noting that the work-list $work_t$ for each thread identified by $t$ is defined as a set containing communities $[t\chi,\ t(\chi+1))\ \cup\ [T\chi + t\chi,\ T\chi + t(\chi+1))\ \cup\ \ldots$, where $\chi$ represents the chunk size and $T$ denotes the total number of threads used. In our implementation, a chunk size of $\chi = 1024$ is utilized for efficient parallel processing. This choice optimizes workload distribution among threads and enhances overall performance during community detection tasks.

\begin{algorithm}[hbtp]
\caption{Split disconnected communities using BFS \cite{sahu2024addressing}.}
\label{alg:splitbfs}
\begin{algorithmic}[1]
\Require{$G(V, E)$: Input graph}
\Require{$C$: Initial community membership/label of each vertex}
\Ensure{$C'$: Updated community membership/label of each vertex}
\Ensure{$f_{if}$: Perform BFS to vertex $j$ if condition satisfied}
\Ensure{$f_{do}$: Perform operation after each vertex is visited}
\Ensure{$vis$: Visited flag for each vertex}
\Ensure{$work_t$: Work-list of current thread}

\Statex

\Function{splitDisconnectedBfs}{$G, C$} \label{alg:splitbfs--begin}
  \State $C' \gets \{\}$ \textbf{;} $vis \gets \{\}$ \label{alg:splitbfs--init-begin}
  \ForAll{$i \in V$ \textbf{in parallel}}
    \State $C'[i] = i$
  \EndFor \label{alg:splitbfs--init-end}
  \ForAll{\textbf{threads}} \label{alg:splitbfs--threads-begin}
    \ForAll{$i \in V$} \label{alg:splitbfs--loop-begin}
      \State $c' \gets C'[i]$ \label{alg:splitbfs--loopinit}
      \If{$c \notin work_t$ \textbf{or} $vis[i]$} \textbf{continue} \label{alg:splitbfs--work}
      \EndIf
      \State $f_{if} \gets (j) \implies C[j] = C[j]$
      \State $f_{do} \gets (j) \implies C'[j] \gets c'$
      \State $bfsVisitForEach(vis, G, i, f_{if}, f_{do})$ \label{alg:splitbfs--bfs}
    \EndFor \label{alg:splitbfs--loop-end}
  \EndFor \label{alg:splitbfs--threads-end}
  \Return{$C'$} \label{alg:splitbfs--return}
\EndFunction \label{alg:splitbfs--end}
\end{algorithmic}
\end{algorithm}

\begin{figure*}[hbtp]
  \centering
  \subfigure[Community $C1$ is internally disconnected]{
    \label{fig:onsplitbfs--1}
    \includegraphics[width=0.31\linewidth]{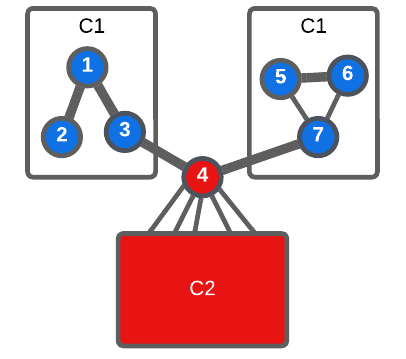}
  }
  \subfigure[Post BFS relabeling from vertex $2$]{
    \label{fig:onsplitbfs--2}
    \includegraphics[width=0.31\linewidth]{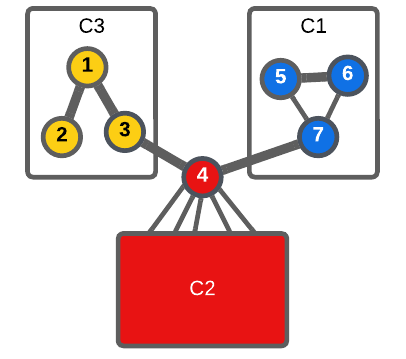}
  }
  \subfigure[Post BFS relabeling from vertex $7$]{
    \label{fig:onsplitbfs--3}
    \includegraphics[width=0.31\linewidth]{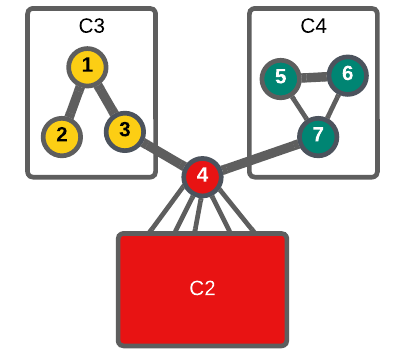}
  } \\[-2ex]
  \caption{An example demonstrating the BFS technique for splitting disconnected communities \cite{sahu2024addressing}. Initially, two communities, $C1$ and $C2$, are depicted, where $C1$ is internally disconnected because vertex $4$ has joined $C2$. The BFS technique randomly selects vertices within each community and assigns the same label to reachable vertices (indicated with a new community ID).}
  \label{fig:onsplitbfs}
\end{figure*}

Figure \ref{fig:onsplitbfs} depicts an example of the Breadth-First Search (BFS) technique. Initially, as shown in Figure \ref{fig:onsplitbfs--1}, two communities, $C1$ and $C2$, are formed after running a few iterations of LPA. Here, $C1$ experiences internal disconnection due to the vertex $4$ selecting $C2$ as its best community, mirroring the scenario illustrated in Figure \ref{fig:onrak--3}. Subsequently, employing the BFS technique, a thread selects a random vertex within community $C1$, such as $2$, and assigns the label $2$ to all vertices reachable within $C1$ from vertex $2$, marking them as visited (Figure \ref{fig:onsplitbfs--2}). Following this, as illustrated in Figure \ref{fig:onsplitbfs--3}, the same thread randomly selects an unvisited vertex within community $C1$, for instance, $7$, and assigns the label $7$ to all vertices reachable within $C1$ from vertex $7$, marking them as visited. A similar procedure is carried out within community $C2$. As a result, all vertices are visited, and the labels assigned to them signify the updated community membership of each vertex without disconnected communities. It's essential to note that each thread possesses a distinct work-list, ensuring that concurrent BFS operations within the same community are avoided.

We also attempted another parallel BFS-based algorithms to split disconnected communities. It begins by initializing the visited flags and BFS frontiers, using OpenMP to parallelize across vertices. Then, for each community, it selects a starting vertex and marks it as visited. Using a BFS approach, it iteratively explores neighbors within each community, updating each unvisited vertex's community label, and marking it as visited if it shares the same community. The algorithm continues this process in a loop, advancing the BFS frontier until no new vertices are marked in a given iteration, at which point a reinitialization is triggered for untouched vertices in each community. This procedure is repeated until all vertices are processed. However, this algorithm offered lower performance than our BFS-based algorithm discussed above.

\subsection{Our GSL-LPA algorithm}

To evaluate the effectiveness of the Split Last (SL) approach in conjunction with the minimum-label-based Label Propagation (LP), minimum-label-based Label Propagation with Pruning (LPP), and Breadth-First Search (BFS) techniques for splitting disconnected communities within the framework of LPA, we employ GVE-LPA \cite{sahu2023gvelouvain}, our parallel implementation of LPA.

\begin{figure}[hbtp]
  \centering
  \subfigure[Relative runtime using \textit{Split Last (SL)} approach, based on three different techniques, for splitting disconnected communities with Parallel LPA]{
    \label{fig:optrak--runtime}
    \includegraphics[width=0.98\linewidth]{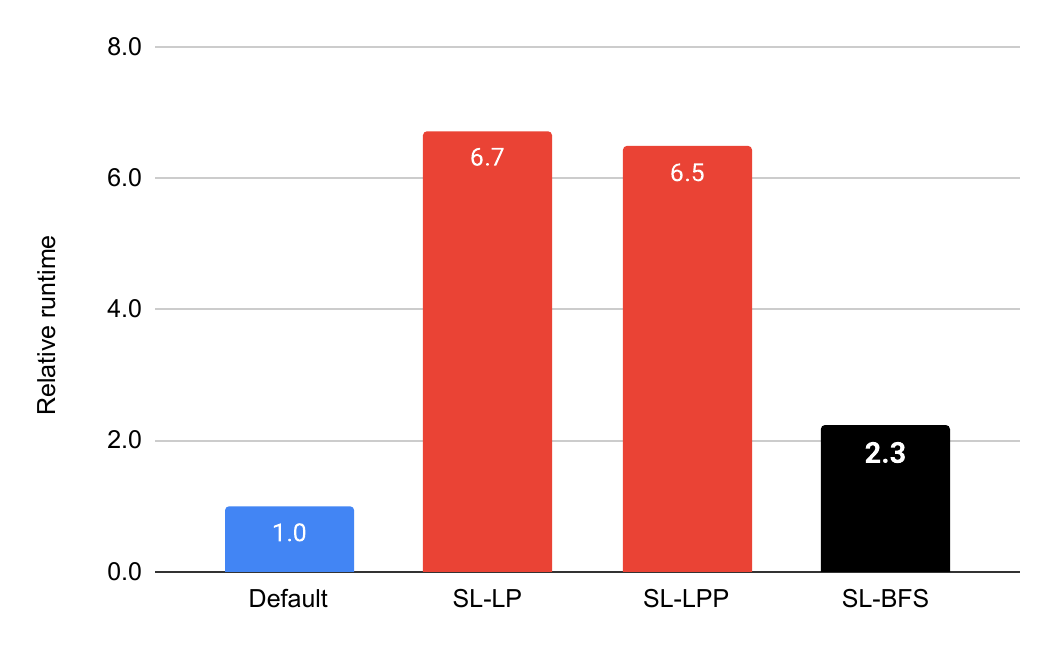}
  } \\[-0ex]
  \subfigure[Modularity of communities obtained using \textit{Split Last (SL)} approach for splitting disconnected communities with Parallel LPA]{
    \label{fig:optrak--modularity}
    \includegraphics[width=0.98\linewidth]{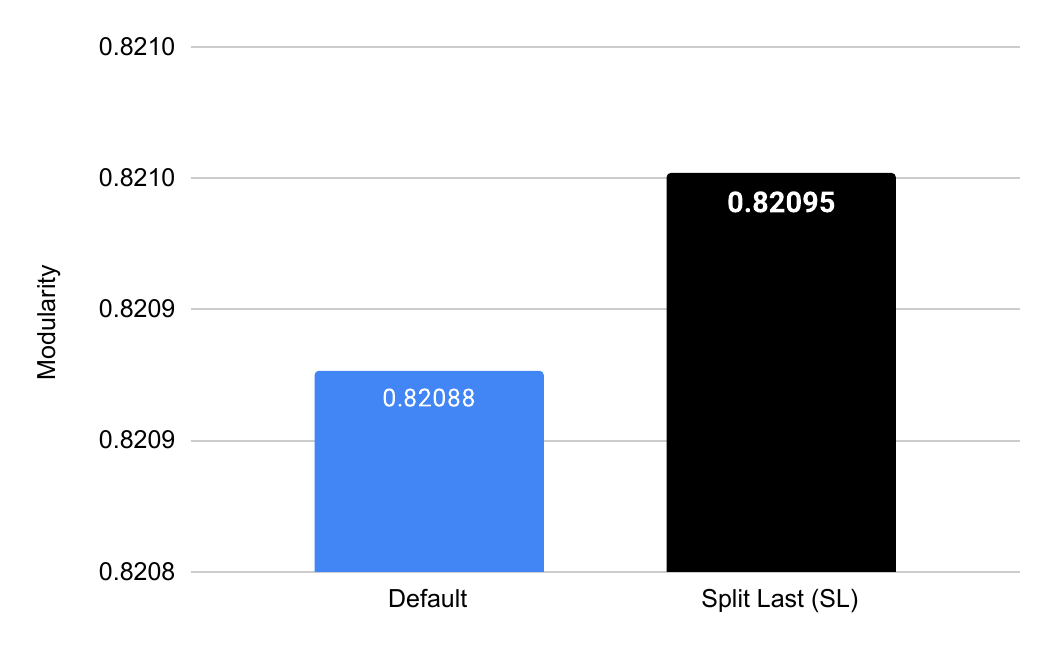}
  } \\[-0ex]
  \subfigure[Fraction of disconnected communities (logarithmic scale) using \textit{Split Last (SL)} approach for splitting disconnected communities with Parallel LPA]{
    \label{fig:optrak--disconnected}
    \includegraphics[width=0.98\linewidth]{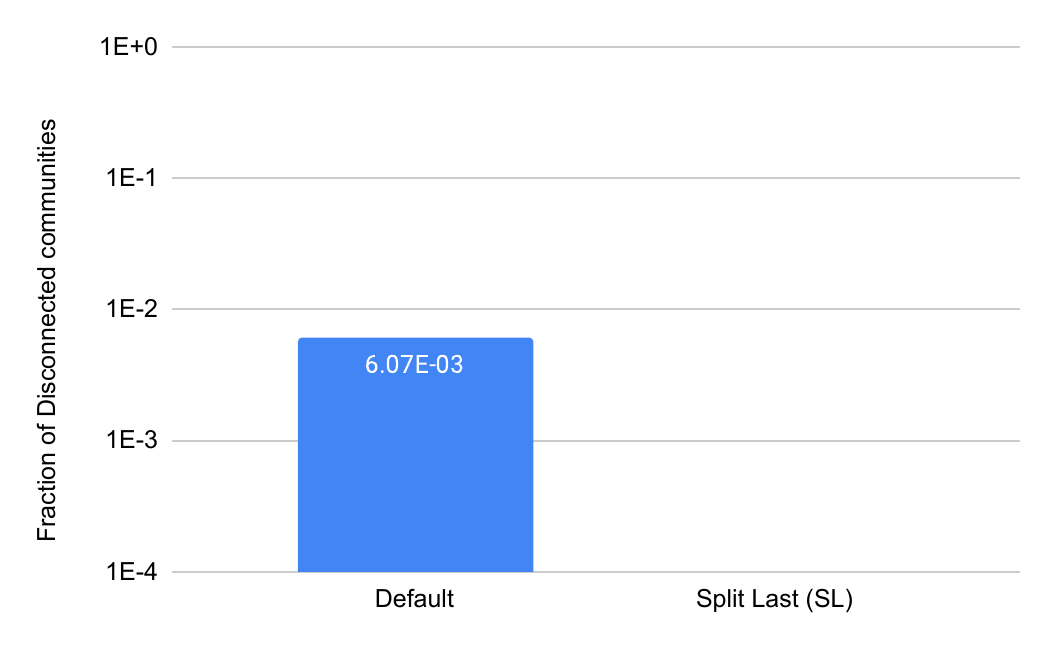}
  } \\[-2ex]
  \caption{Mean relative runtime, modularity, and fraction of disconnected communities (log-scale) using the \textit{Split Last (SL)} approach for addressing disconnected communities with Parallel LPA \cite{sahu2023gvelpa} across all graphs in the dataset. The \textit{SL} approach utilizes minimum label-based \textit{Label Propagation (LP)}, \textit{Label Propagation with Pruning (LPP)}, or \textit{Breadth First Search (BFS)} techniques for splitting disconnected communities.}
  \label{fig:optrak}
\end{figure}

\subsubsection{Determining suitable technique for splitting disconnected communities}

To evaluate the optimal technique for partitioning disconnected communities, we investigate the SL approach with LP, LPP, and BFS techniques. Figures \ref{fig:optrak--runtime}, \ref{fig:optrak--modularity}, and \ref{fig:optrak--disconnected} display the mean relative runtime, modularity, and fractions of disconnected communities for SL-LP, SL-LPP, SL-BFS, and the default approaches. The SL approach yields non-disconnected communities, as depicted in Figure \ref{fig:optrak--disconnected}. Additionally, Figure \ref{fig:optrak--modularity} shows that the modularity of communities obtained through the SL approach exceeds that of the default approach. Figure \ref{fig:optrak--runtime} illustrates that SL-BFS, i.e., the SL approach with BFS technique, exhibits superior performance. Thus, employing BFS for splitting disconnected communities as a post-processing step (SL) of LPA is preferable.

\begin{algorithm}[hbtp]
\caption{GSL-LPA: Our Parallel LPA which identifies communities that are not internally disconnected.}
\label{alg:rak}
\begin{algorithmic}[1]
\Require{$G(V, E)$: Input graph}
\Ensure{$C$: Community membership of each vertex}
\Ensure{$H_t$: Collision-free per-thread hashtable}
\Ensure{$l_i$: Number of iterations performed}
\Ensure{$\tau$: Per iteration tolerance}

\Statex

\Function{lpa}{$G$} \label{alg:rak--main-begin}
  \State Vertex membership: $C \gets [0 .. |V|)$ \label{alg:rak--init-begin}
  \State Mark all vertices in $G$ as unprocessed \label{alg:rak--init-end}
  \ForAll{$l_i \in [0 .. \text{\small{MAX\_ITERATIONS}})$} \label{alg:rak--iters-begin}
    \State $\Delta N \gets lpaMove(G, C)$ \label{alg:rak--propagate}
    \If{$\Delta N/N \le \tau$} \textbf{break} \Comment{Converged?} \label{alg:rak--converged}
    \EndIf
  \EndFor \label{alg:rak--iters-end}
  \State $C \gets splitDisconnectedBfs(G, C)$ \Comment{Algorithm \ref{alg:splitbfs}} \label{alg:rak--split}
  \Return{$C$} \label{alg:rak--main-return}
\EndFunction \label{alg:rak--main-end}

\Statex

\Function{lpaMove}{$G, C$} \label{alg:rak--move-begin}
  \State Changed vertices: $\Delta N \gets 0$
  \ForAll{unprocessed $i \in V$ \textbf{in parallel}}
    \State Mark $i$ as processed (prune) \label{alg:rak--prune}
    \State $H_t \gets scanCommunities(\{\}, G, C, i)$ \label{alg:rak--scan}
    \State $\rhd$ Use $H_t$ to choose the most weighted label
    \State $c^* \gets$ Most weighted label to $i$ in $G$ \label{alg:rak--best-community}
    \If{$c^* = C[i]$} \textbf{continue} \label{alg:rak--best-community-same}
    \EndIf
    \State $C[i] \gets c^*$ \textbf{;} $\Delta N \gets \Delta N + 1$ \label{alg:rak--perform-move}
    \State Mark neighbors of $i$ as unprocessed \label{alg:rak--remark}
  \EndFor
\Return{$\Delta N$} \label{alg:rak--move-return}
\EndFunction \label{alg:rak--move-end}

\Statex

\Function{scanCommunities}{$H_t, G, C, i$} \label{alg:rak--scan-begin}
  \ForAll{$(j, w) \in G.edges(i)$}
    \If{$i \neq j$} $H_t[C[j]] \gets H_t[C[j]] + w$
    \EndIf
  \EndFor
  \Return{$H_t$}
\EndFunction \label{alg:rak--scan-end}
\end{algorithmic}
\end{algorithm}

\subsubsection{Explanation of the algorithm}

We denote GVE-LPA utilizing the Split Last (SL) approach with Breadth-First Search (SL-BFS) technique to deal with the internally-disconnected communities as \textit{GSL-LPA}. The core procedure of GSL-LPA, encapsulated in the \texttt{lpa()} function, is presented in lines \ref{alg:rak--main-begin}-\ref{alg:rak--main-end} of Algorithm \ref{alg:rak}. This function takes a graph $G(V, E)$ as input and produces the community membership (or label) $C$ for each vertex in the graph, ensuring that none of the returned communities are internally disconnected.

In lines \ref{alg:rak--init-begin}-\ref{alg:rak--init-end}, we start by initializing the labels $C$ for each vertex in $G$, and marking all vertices in the graph as unprocessed. We then iterate to propagate labels based on the weighted influence of neighboring vertices. We limit the number of iterations to $MAX\_ITERATIONS$ (lines \ref{alg:rak--iters-begin}-\ref{alg:rak--iters-end}). Within each iteration, we utilize the \texttt{lpaMove()} function for label propagation and track the count of nodes with updated labels $\Delta N$ (line \ref{alg:rak--propagate}). If the ratio of $\Delta N$ to the total number of nodes $N$ falls within the specified iteration tolerance $\tau$, convergence is achieved, and the loop terminates (line \ref{alg:rak--converged}). After completing all iterations, the algorithm enters the splitting phase to separate internally-disconnected communities in $C$. This is accomplished using the parallel BFS technique in line \ref{alg:rak--split}, employing the \texttt{splitCommunitiesBfs()} function (Algorithm \ref{alg:splitbfs}). Finally, we return the updated labels $C$ (line \ref{alg:rak--main-return}).

In lines \ref{alg:rak--move-begin}-\ref{alg:rak--move-end}, the function \texttt{lpaMove()} operates by concurrently iterating over unprocessed vertices. For each unprocessed vertex $i$ in the graph $G$, it marks $i$ as processed (vertex pruning optimization, line \ref{alg:rak--prune}), computes the total edge weight of connected labels in a per-thread hashtable $H_t$ using the \texttt{scanCommunities()} function (line \ref{alg:rak--scan}), and selects the most weighted label $c^*$ (line \ref{alg:rak--best-community}). If $c^*$ differs from the current label of $i$, the label of $i$ is updated, the count of changed vertices $\Delta N$ is incremented, and the neighbors of $i$ are marked as unprocessed for the next iteration (lines \ref{alg:rak--perform-move}-\ref{alg:rak--remark}). Upon having processed all the vertices, the function returns the total number of vertices with updated labels $\Delta N$ (line \ref{alg:rak--move-return}). The \texttt{scanCommunities()} function, given in lines \ref{alg:rak--scan-begin}-\ref{alg:rak--scan-end}, iterates over the neighbors of the current vertex $i$, excluding itself, and computes the total edge weight linked to each label in the hashtable $H_t$.

\section{Evaluation}
\label{sec:evaluation}
\subsection{Experimental Setup}
\label{sec:setup}

\subsubsection{System used}

We employ a server consisting of two Intel Xeon Gold 6226R processors, each featuring $16$ cores running at $2.90$ GHz. Each core includes a $1$ MB L1 cache, a $16$ MB L2 cache, and shares a $22$ MB L3 cache. The system is equipped with $376$ GB of RAM and runs CentOS Stream 8.

\subsubsection{Configuration}

We utilize 32-bit integers for representing vertex IDs and 32-bit floats for edge weights, while computations and hashtable values are based on 64-bit floats. We use $64$ threads by default, which to correspond to the available cores on the system. Compilation is conducted using GCC 8.5 and OpenMP 4.5.

\subsubsection{Dataset}
\label{sec:dataset}

The graphs employed in our experiments are detailed in Table \ref{tab:dataset}, obtained from the SuiteSparse Matrix Collection \cite{suite19}. These graphs encompass a range of $3.07$ to $214$ million vertices and $25.4$ million to $3.80$ billion edges. We make sure that the edges are both undirected and weighted, with a default weight set to $1$.

\begin{table}[hbtp]
  \centering
  \caption{List of $13$ graphs sourced from the SuiteSparse Matrix Collection \cite{suite19} (directed graphs are marked with $*$). Here, $|V|$ is the number of vertices, $|E|$ is the number of edges (after adding reverse edges), $D_{avg}$ is the average degree, and $|\Gamma|$ is the number of communities identified with GSL-LPA.}
  \label{tab:dataset}
  \begin{tabular}{|c||c|c|c|c|}
    \toprule
    \textbf{Graph} &
    \textbf{\textbf{$|V|$}} &
    \textbf{\textbf{$|E|$}} &
    \textbf{\textbf{$D_{avg}$}} &
    \textbf{\textbf{$|\Gamma|$}} \\
    \midrule
    \multicolumn{5}{|c|}{\textbf{Web Graphs (LAW)}} \\ \hline
    indochina-2004$^*$ & 7.41M & 341M & 41.0 & 160K \\ \hline
    uk-2002$^*$ & 18.5M & 567M & 16.1 & 414K \\ \hline
    arabic-2005$^*$ & 22.7M & 1.21B & 28.2 & 268K \\ \hline
    uk-2005$^*$ & 39.5M & 1.73B & 23.7 & 744K \\ \hline
    webbase-2001$^*$ & 118M & 1.89B & 8.6 & 6.70M \\ \hline
    it-2004$^*$ & 41.3M & 2.19B & 27.9 & 694K \\ \hline
    sk-2005$^*$ & 50.6M & 3.80B & 38.5 & 433K \\ \hline
    \multicolumn{5}{|c|}{\textbf{Social Networks (SNAP)}} \\ \hline
    com-LiveJournal & 4.00M & 69.4M & 17.4 & 6.91K \\ \hline
    com-Orkut & 3.07M & 234M & 76.2 & 192 \\ \hline
    \multicolumn{5}{|c|}{\textbf{Road Networks (DIMACS10)}} \\ \hline
    asia\_osm & 12.0M & 25.4M & 2.1 & 278K \\ \hline
    europe\_osm & 50.9M & 108M & 2.1 & 1.52M \\ \hline
    \multicolumn{5}{|c|}{\textbf{Protein k-mer Graphs (GenBank)}} \\ \hline
    kmer\_A2a & 171M & 361M & 2.1 & 40.1M \\ \hline
    kmer\_V1r & 214M & 465M & 2.2 & 48.7M \\ \hline
  \bottomrule
  \end{tabular}
\end{table}

\begin{figure*}[hbtp]
  \centering
  \subfigure[Runtime in seconds (logarithmic scale) with \textit{FLPA}, \textit{igraph LPA}, \textit{NetworKit LPA}, and \textit{GSL-LPA}]{
    \label{fig:cmprak--runtime}
    \includegraphics[width=0.98\linewidth]{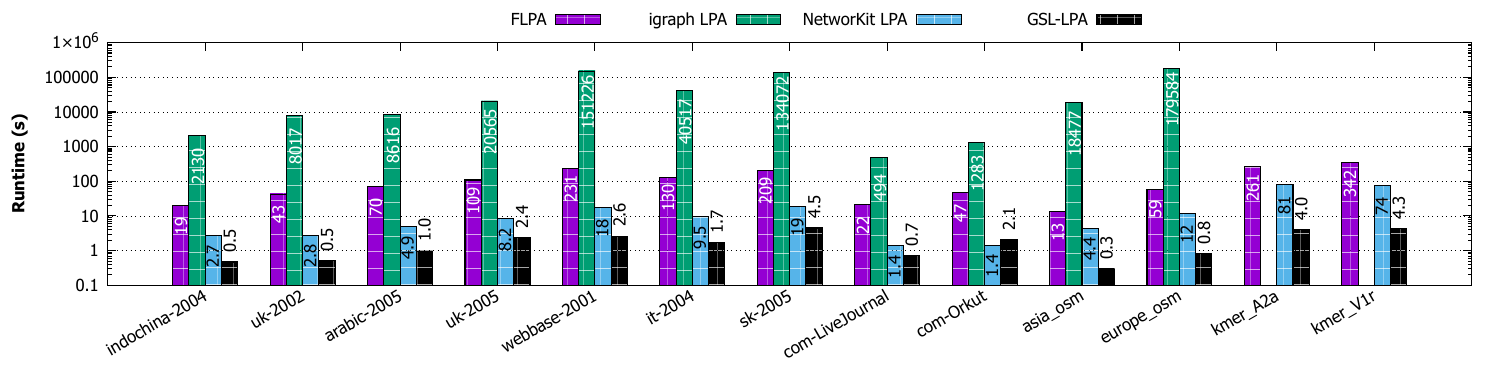}
  } \\[-0ex]
  \subfigure[Speedup (logarithmic scale) of \textit{GSL-LPA} with respect to \textit{FLPA}, \textit{igraph LPA}, \textit{NetworKit LPA}.]{
    \label{fig:cmprak--speedup}
    \includegraphics[width=0.98\linewidth]{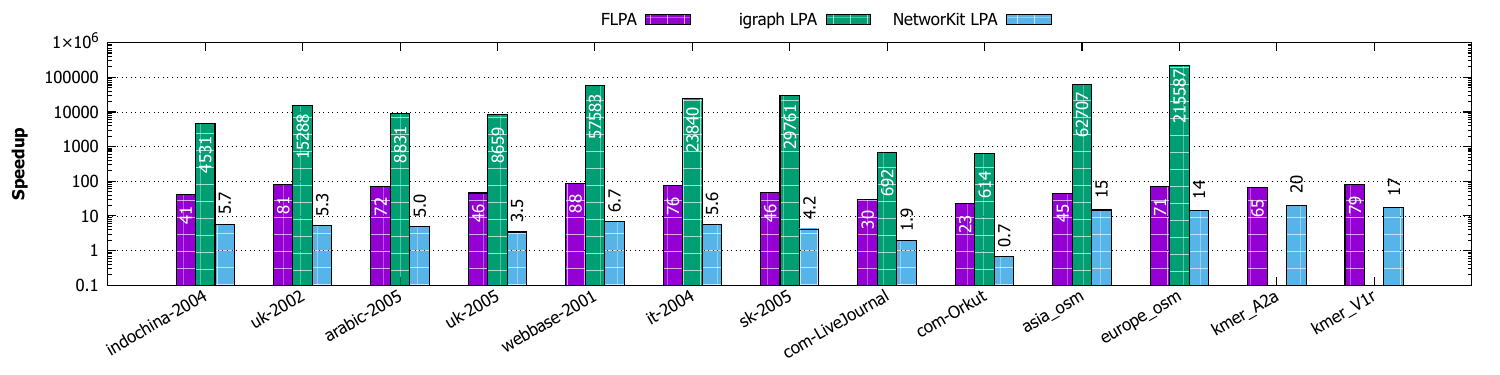}
  } \\[-0ex]
  \subfigure[Modularity of communities obtained with \textit{FLPA}, \textit{igraph LPA}, \textit{NetworKit LPA}, and \textit{GSL-LPA}.]{
    \label{fig:cmprak--modularity}
    \includegraphics[width=0.98\linewidth]{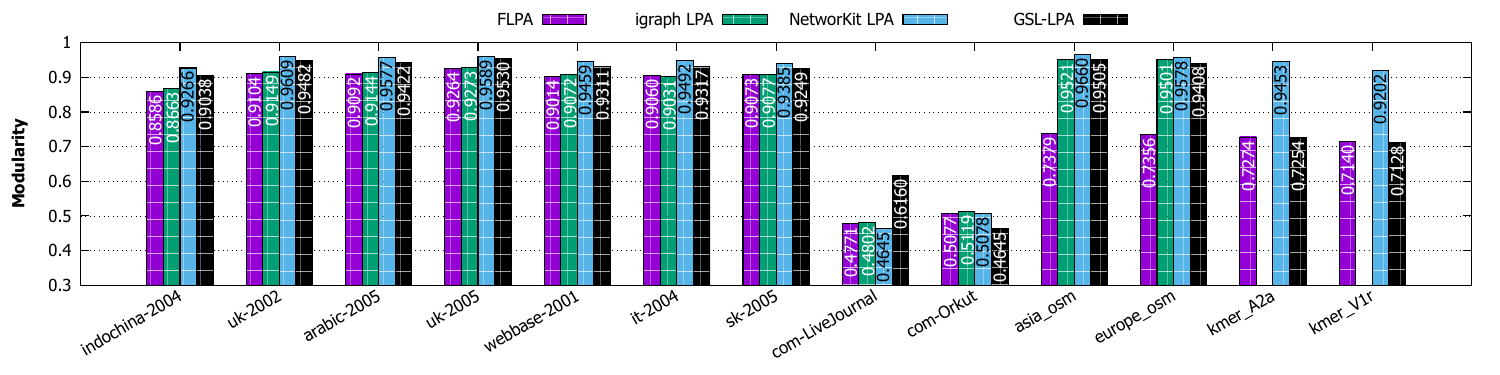}
  } \\[-0ex]
  \subfigure[Fraction of disconnected communities (logarithmic scale) with \textit{FLPA}, \textit{igraph LPA}, \textit{NetworKit LPA}, and \textit{GSL-LPA}.]{
    \label{fig:cmprak--disconnected}
    \includegraphics[width=0.98\linewidth]{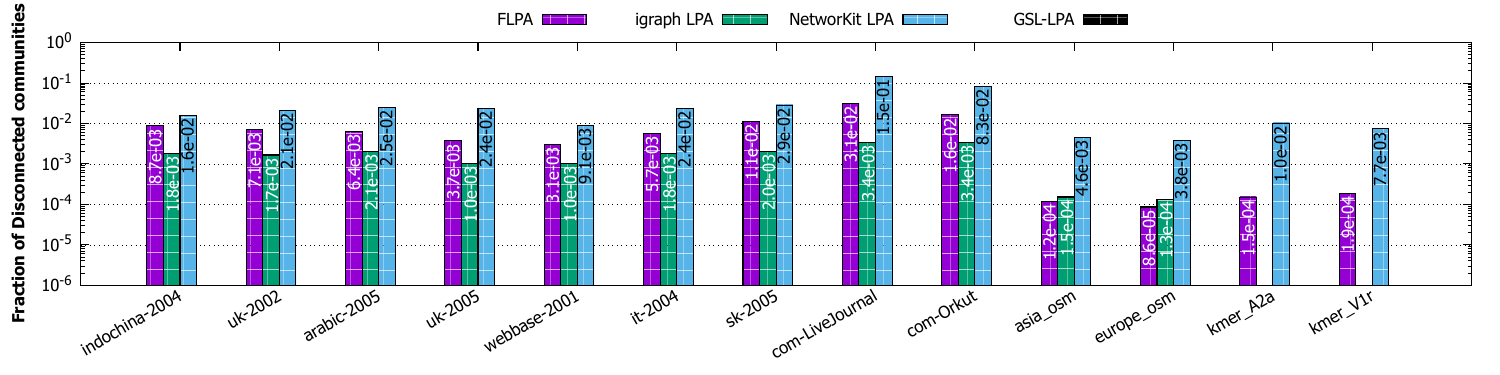}
  } \\[-2ex]
  \caption{Runtime in seconds (log-scale), speedup (log-scale), modularity, and fraction of disconnected communities (log-scale) compared across \textit{FLPA}, \textit{igraph LPA}, \textit{NetworKit LPA}, and \textit{GSL-LPA} for each graph in the dataset. \textit{igraph LPA} fails to execute on \textit{kmer\_A2a} and \textit{kmer\_V1r} graphs, and thus its results are excluded.}
  \label{fig:cmprak}
\end{figure*}

\subsection{Comparing Performance of GSL-LPA}

We now compare the performance of GSL-LPA with igraph LPA \cite{csardi2006igraph}, FLPA \cite{traag2023large}, and NetworKit LPA \cite{staudt2016networkit}. Both igraph LPA and FLPA are implemented sequentially, while NetworKit LPA is parallelized. For FLPA, we checkout the branch featuring the modified version of \texttt{igraph\_community\_label\_propagation()} function, and update the label propagation example in C to use the \texttt{IGRAPH\_LPA\_FAST} variant, and enable loading the input graph from a file. The runtime of \texttt{igraph\_community\_label\_propagation()} is measured using \texttt{gettimeofday()}. For igraph LPA, we adopt a similar methodology as FLPA, but utilize the code from the \texttt{master} branch of igraph. For NetworKit, we use a Python script to run \texttt{PLP} (Parallel Label Propagation) algorithm, and record the runtime with \texttt{getTiming()}. The runtime of NetworKit LPA is measured five times for each graph to obtain an average. Due to extensive duration of igraph LPA runs, we perform only a single run per graph. Additionally, we record the modularity of communities identified by each implementation.

Figure \ref{fig:cmprak--runtime} depicts the runtimes of FLPA, igraph LPA, NetworKit LPA, and GVE-LPA for each graph in the dataset. Meanwhile, Figure \ref{fig:cmprak--speedup} illustrates the speedup of GSL-LPA compared to each of the aforementioned implementations of LPA. igraph LPA encounters runtime issues on protein k-mer graphs (\textit{kmer\_A2a} and \textit{kmer\_V1r}), hence its results for these graphs are omitted.

GSL-LPA demonstrates average speedups of $55\times$, $10,300\times$, and $5.8\times$ over FLPA, igraph LPA, and NetworKit LPA respectively. Specifically, on the \textit{sk-2005} graph, GSL-LPA identifies communities within $4.5$ seconds, achieving a processing rate of $844$ million edges/s. Furthermore, Figure \ref{fig:cmprak--modularity} presents the modularity of communities obtained by each implementation. On average, GSL-LPA achieves modularity that is $7.1\%$ and $0.7\%$ higher than FLPA (particularly on road networks and protein k-mer graphs) and igraph LPA respectively. Additionally, it records a $3.6\%$ lower modularity compared to NetworKit LPA (especially on protein k-mer graphs). Finally, Figure \ref{fig:cmprak--disconnected} illustrates the fraction of disconnected communities obtained by each implementation. The absence of bars indicates the absence of disconnected communities. Communities identified by the GSL-LPA exhibit no disconnected communities. However, on average, FLPA, igraph LPA, and NetworKit LPA exhibit fractions of disconnected communities amounting to $2.3\times10^{-2}$, $1.2\times10^{-3}$, and $1.8\times10^{-2}$, respectively, particularly noticeable on web graphs and social networks.\ignore{Thus, GSL-LPA effectively tackles the issue of disconnected communities, while being significantly faster than existing alternatives, and attaining similar modularity scores.} Figure \ref{fig:cmpduo} depicts the comparison of GVE-LPA and GSL-LPA. This comparison is explained in detail in Section \ref{sec:comparison-extra}.

\begin{figure}[hbtp]
  \centering
  \subfigure{
    \label{fig:splitrak--phase}
    \includegraphics[width=0.98\linewidth]{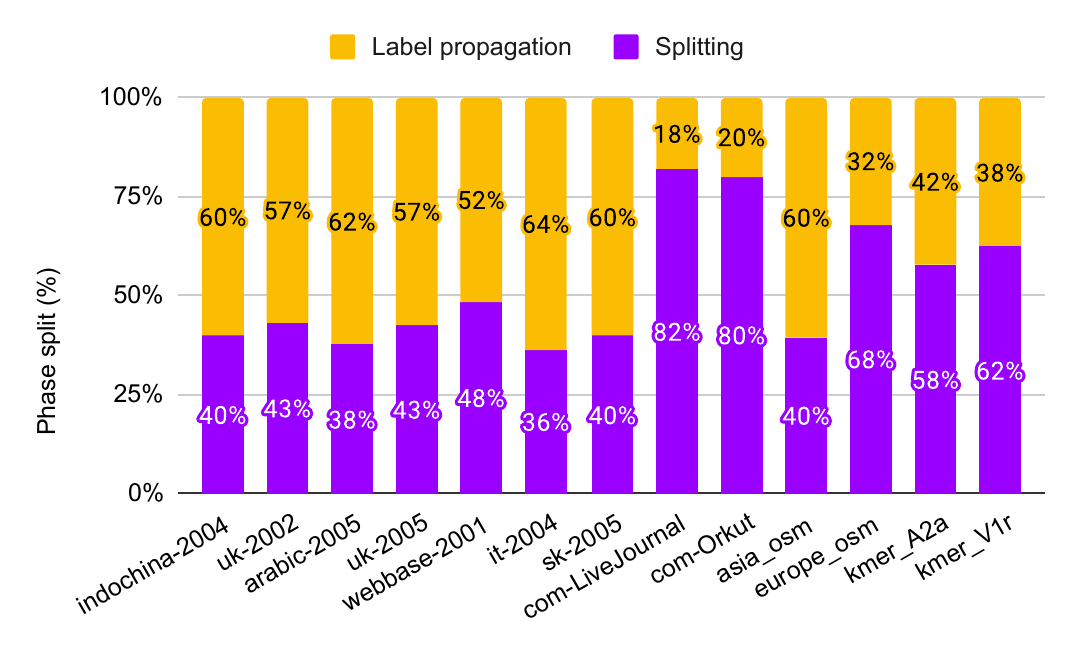}
  } \\[-2ex]
  \caption{Phase split of \textit{GSL-LPA} for each graph in the dataset.}
  \label{fig:splitrak}
\end{figure}

\begin{figure}[hbtp]
  \centering
  \subfigure{
    \label{fig:scaling--rak}
    \includegraphics[width=0.98\linewidth]{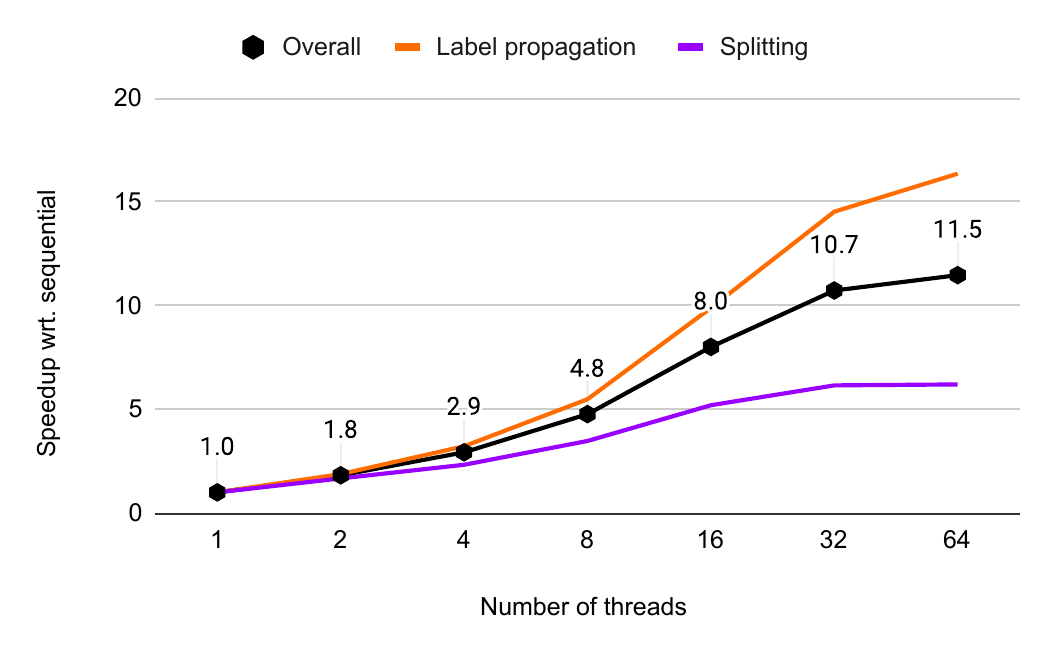}
  } \\[-2ex]
  \caption{Overall speedup of \textit{GSL-LPA}, including its label propagation and splitting phases, with increasing number of threads (in powers of 2).}
  \label{fig:scaling}
\end{figure}

\subsection{Performance Analysis of GSL-LPA}
\label{sec:analysis}

Next, we proceed with an analysis of GSL-LPA's performance. The phase-wise breakdown of GSL-Leiden is depicted in Figure \ref{fig:splitrak}. This illustration shows that GSL-LPA allocates a significant portion of its runtime to the splitting phase, particularly noticeable on social networks, whereas it predominantly emphasizes the label propagation phase on web graphs. On average, GSL-LPA dedicates $47\%$ of its runtime to the label propagation phase and $53\%$ to the splitting phase. We plan to address the optimization of the splitting phase, i.e., Algorithm \ref{alg:splitbfs}, in future work.

\subsection{Strong Scaling of GSL-LPA}

Finally, we evaluate the strong scaling performance of GSL-LPA. This involves varying the number of threads from $1$ to $64$ in powers of $2$, for each input graph, and measuring the time taken for GSL-LPA to identify communities. Figure \ref{fig:scaling} illustrates the average strong scaling of GSL-LPA, including the scaling of its individual phases: the label propagation phase and the splitting phase. At 32 threads, GSL-LPA achieves a speedup of $10.7\times$ compared to single-threaded execution, indicating a performance increase of $1.6\times$ for each doubling of threads. The scalability is constrained by the splitting phase, which only provides a speedup of $6.2\times$ with 32 threads, while the label propagation phase offers a speedup of $14.5\times$ with the same number of threads. At 64 threads, GSL-LPA is affected by NUMA effects and offers a speedup of $11.5\times$.

\section{Conclusion}
\label{sec:conclusion}
In this report, we explored the issue of internally-disconnected communities arising from the Label Propagation Algorithm (LPA). Our experimental findings showed that FLPA \cite{traag2023large}, igraph LPA \cite{csardi2006igraph}, and NetworKit LPA \cite{staudt2016networkit} identified communities with this issue in up to $3.1\%$, $0.3\%$, and $14.5\%$ of cases respectively. To address this, we introduced GSL-LPA, derived from our parallelization of LPA (GVE-LPA) \cite{sahu2023gvelpa}, which mitigates this problem by splitting disconnected communities as a post-processing step, akin to previous approaches. Our experiments conducted on a system equipped with two 16-core Intel Xeon Gold 6226R processors demonstrated that GSL-LPA not only resolves this issue but also outperforms FLPA, igraph LPA, and NetworKit LPA by $55\times$, $10,300\times$, and $5.8\times$ respectively. Moreover, GSL-LPA achieves a processing rate of $844 M$ edges/s on a $3.8 B$ edge graph. Additionally, GSL-LPA exhibits performance improvement at a rate of $1.6\times$ for every doubling of threads.

\begin{acks}
I would like to thank Prof. Kishore Kothapalli and Prof. Dip Sankar Banerjee for their support.
\end{acks}

\bibliographystyle{ACM-Reference-Format}
\bibliography{main}


\begin{thebibliography}{59}


\ifx \showCODEN    \undefined \def \showCODEN     #1{\unskip}     \fi
\ifx \showDOI      \undefined \def \showDOI       #1{#1}\fi
\ifx \showISBNx    \undefined \def \showISBNx     #1{\unskip}     \fi
\ifx \showISBNxiii \undefined \def \showISBNxiii  #1{\unskip}     \fi
\ifx \showISSN     \undefined \def \showISSN      #1{\unskip}     \fi
\ifx \showLCCN     \undefined \def \showLCCN      #1{\unskip}     \fi
\ifx \shownote     \undefined \def \shownote      #1{#1}          \fi
\ifx \showarticletitle \undefined \def \showarticletitle #1{#1}   \fi
\ifx \showURL      \undefined \def \showURL       {\relax}        \fi
\providecommand\bibfield[2]{#2}
\providecommand\bibinfo[2]{#2}
\providecommand\natexlab[1]{#1}
\providecommand\showeprint[2][]{arXiv:#2}

\bibitem[Abbe(2018)]%
        {abbe2018community}
\bibfield{author}{\bibinfo{person}{Emmanuel Abbe}.} \bibinfo{year}{2018}\natexlab{}.
\newblock \showarticletitle{Community detection and stochastic block models: recent developments}.
\newblock \bibinfo{journal}{\emph{Journal of Machine Learning Research}} \bibinfo{volume}{18}, \bibinfo{number}{177} (\bibinfo{year}{2018}), \bibinfo{pages}{1--86}.
\newblock


\bibitem[Ball et~al\mbox{.}(2011)]%
        {ball2011efficient}
\bibfield{author}{\bibinfo{person}{B. Ball}, \bibinfo{person}{B. Karrer}, {and} \bibinfo{person}{M.~EJ. Newman}.} \bibinfo{year}{2011}\natexlab{}.
\newblock \showarticletitle{Efficient and principled method for detecting communities in networks}.
\newblock \bibinfo{journal}{\emph{Physical Review E}} \bibinfo{volume}{84}, \bibinfo{number}{3} (\bibinfo{year}{2011}), \bibinfo{pages}{036103}.
\newblock


\bibitem[Bechtel et~al\mbox{.}(2005)]%
        {bechtel2005lung}
\bibfield{author}{\bibinfo{person}{Joel~J Bechtel}, \bibinfo{person}{William~A Kelley}, \bibinfo{person}{Teresa~A Coons}, \bibinfo{person}{M~Gerry Klein}, \bibinfo{person}{Daniel~D Slagel}, {and} \bibinfo{person}{Thomas~L Petty}.} \bibinfo{year}{2005}\natexlab{}.
\newblock \showarticletitle{Lung cancer detection in patients with airflow obstruction identified in a primary care outpatient practice}.
\newblock \bibinfo{journal}{\emph{Chest}} \bibinfo{volume}{127}, \bibinfo{number}{4} (\bibinfo{year}{2005}), \bibinfo{pages}{1140--1145}.
\newblock


\bibitem[Berahmand and Bouyer(2018)]%
        {com-berahmand18}
\bibfield{author}{\bibinfo{person}{K. Berahmand} {and} \bibinfo{person}{A. Bouyer}.} \bibinfo{year}{2018}\natexlab{}.
\newblock \showarticletitle{{LP-LPA: A link influence-based label propagation algorithm for discovering community structures in networks}}.
\newblock \bibinfo{journal}{\emph{International Journal of Modern Physics B}} \bibinfo{volume}{32}, \bibinfo{number}{06} (\bibinfo{date}{10 mar} \bibinfo{year}{2018}), \bibinfo{pages}{1850062}.
\newblock
\showISSN{0217-9792}


\bibitem[Blondel et~al\mbox{.}(2008)]%
        {com-blondel08}
\bibfield{author}{\bibinfo{person}{V. Blondel}, \bibinfo{person}{J. Guillaume}, \bibinfo{person}{R. Lambiotte}, {and} \bibinfo{person}{E. Lefebvre}.} \bibinfo{year}{2008}\natexlab{}.
\newblock \showarticletitle{{Fast unfolding of communities in large networks}}.
\newblock \bibinfo{journal}{\emph{Journal of Statistical Mechanics: Theory and Experiment}} \bibinfo{volume}{2008}, \bibinfo{number}{10} (\bibinfo{date}{Oct} \bibinfo{year}{2008}), \bibinfo{pages}{P10008}.
\newblock


\bibitem[Brandes et~al\mbox{.}(2007)]%
        {com-brandes07}
\bibfield{author}{\bibinfo{person}{U. Brandes}, \bibinfo{person}{D. Delling}, \bibinfo{person}{M. Gaertler}, \bibinfo{person}{R. Gorke}, \bibinfo{person}{M. Hoefer}, \bibinfo{person}{Z. Nikoloski}, {and} \bibinfo{person}{D. Wagner}.} \bibinfo{year}{2007}\natexlab{}.
\newblock \showarticletitle{{On modularity clustering}}.
\newblock \bibinfo{journal}{\emph{IEEE transactions on knowledge and data engineering}} \bibinfo{volume}{20}, \bibinfo{number}{2} (\bibinfo{year}{2007}), \bibinfo{pages}{172--188}.
\newblock


\bibitem[Clauset et~al\mbox{.}(2004)]%
        {clauset2004finding}
\bibfield{author}{\bibinfo{person}{Aaron Clauset}, \bibinfo{person}{Mark~EJ Newman}, {and} \bibinfo{person}{Cristopher Moore}.} \bibinfo{year}{2004}\natexlab{}.
\newblock \showarticletitle{Finding community structure in very large networks}.
\newblock \bibinfo{journal}{\emph{Physical review E}} \bibinfo{volume}{70}, \bibinfo{number}{6} (\bibinfo{year}{2004}), \bibinfo{pages}{066111}.
\newblock


\bibitem[Cordasco and Gargano(2012)]%
        {cordasco2012label}
\bibfield{author}{\bibinfo{person}{Gennaro Cordasco} {and} \bibinfo{person}{Luisa Gargano}.} \bibinfo{year}{2012}\natexlab{}.
\newblock \showarticletitle{Label propagation algorithm: a semi-synchronous approach}.
\newblock \bibinfo{journal}{\emph{International Journal of Social Network Mining}} \bibinfo{volume}{1}, \bibinfo{number}{1} (\bibinfo{year}{2012}), \bibinfo{pages}{3--26}.
\newblock


\bibitem[Coscia et~al\mbox{.}(2011)]%
        {coscia2011classification}
\bibfield{author}{\bibinfo{person}{Michele Coscia}, \bibinfo{person}{Fosca Giannotti}, {and} \bibinfo{person}{Dino Pedreschi}.} \bibinfo{year}{2011}\natexlab{}.
\newblock \showarticletitle{A classification for community discovery methods in complex networks}.
\newblock \bibinfo{journal}{\emph{Statistical Analysis and Data Mining: The ASA Data Science Journal}} \bibinfo{volume}{4}, \bibinfo{number}{5} (\bibinfo{year}{2011}), \bibinfo{pages}{512--546}.
\newblock


\bibitem[Csardi et~al\mbox{.}(2006)]%
        {csardi2006igraph}
\bibfield{author}{\bibinfo{person}{G. Csardi}, \bibinfo{person}{T. Nepusz}, {et~al\mbox{.}}} \bibinfo{year}{2006}\natexlab{}.
\newblock \showarticletitle{The igraph software package for complex network research}.
\newblock \bibinfo{journal}{\emph{InterJournal, complex systems}} \bibinfo{volume}{1695}, \bibinfo{number}{5} (\bibinfo{year}{2006}), \bibinfo{pages}{1--9}.
\newblock


\bibitem[Duch and Arenas(2005)]%
        {duch2005community}
\bibfield{author}{\bibinfo{person}{Jordi Duch} {and} \bibinfo{person}{Alex Arenas}.} \bibinfo{year}{2005}\natexlab{}.
\newblock \showarticletitle{Community detection in complex networks using extremal optimization}.
\newblock \bibinfo{journal}{\emph{Physical review E}} \bibinfo{volume}{72}, \bibinfo{number}{2} (\bibinfo{year}{2005}), \bibinfo{pages}{027104}.
\newblock


\bibitem[Fortunato(2010)]%
        {com-fortunato10}
\bibfield{author}{\bibinfo{person}{S. Fortunato}.} \bibinfo{year}{2010}\natexlab{}.
\newblock \showarticletitle{{Community detection in graphs}}.
\newblock \bibinfo{journal}{\emph{Physics reports}} \bibinfo{volume}{486}, \bibinfo{number}{3-5} (\bibinfo{year}{2010}), \bibinfo{pages}{75--174}.
\newblock


\bibitem[Garza and Schaeffer(2019)]%
        {garza2019community}
\bibfield{author}{\bibinfo{person}{Sara~E Garza} {and} \bibinfo{person}{Satu~Elisa Schaeffer}.} \bibinfo{year}{2019}\natexlab{}.
\newblock \showarticletitle{Community detection with the label propagation algorithm: a survey}.
\newblock \bibinfo{journal}{\emph{Physica A: Statistical Mechanics and its Applications}}  \bibinfo{volume}{534} (\bibinfo{year}{2019}), \bibinfo{pages}{122058}.
\newblock


\bibitem[Gregory(2010)]%
        {com-gregory10}
\bibfield{author}{\bibinfo{person}{S. Gregory}.} \bibinfo{year}{2010}\natexlab{}.
\newblock \showarticletitle{{Finding overlapping communities in networks by label propagation}}.
\newblock \bibinfo{journal}{\emph{New Journal of Physics}}  \bibinfo{volume}{12} (\bibinfo{date}{10} \bibinfo{year}{2010}), \bibinfo{pages}{103018}.
\newblock
Issue 10.


\bibitem[Gui et~al\mbox{.}(2018)]%
        {gui2018community}
\bibfield{author}{\bibinfo{person}{Qiong Gui}, \bibinfo{person}{Rui Deng}, \bibinfo{person}{Pengfei Xue}, {and} \bibinfo{person}{Xiaohui Cheng}.} \bibinfo{year}{2018}\natexlab{}.
\newblock \showarticletitle{A community discovery algorithm based on boundary nodes and label propagation}.
\newblock \bibinfo{journal}{\emph{Pattern Recognition Letters}}  \bibinfo{volume}{109} (\bibinfo{year}{2018}), \bibinfo{pages}{103--109}.
\newblock


\bibitem[Guimer{\`a} et~al\mbox{.}(2010)]%
        {guimera2010origin}
\bibfield{author}{\bibinfo{person}{Roger Guimer{\`a}}, \bibinfo{person}{DB Stouffer}, \bibinfo{person}{Marta Sales-Pardo}, \bibinfo{person}{EA Leicht}, \bibinfo{person}{MEJ Newman}, {and} \bibinfo{person}{Luis~AN Amaral}.} \bibinfo{year}{2010}\natexlab{}.
\newblock \showarticletitle{Origin of compartmentalization in food webs}.
\newblock \bibinfo{journal}{\emph{Ecology}} \bibinfo{volume}{91}, \bibinfo{number}{10} (\bibinfo{year}{2010}), \bibinfo{pages}{2941--2951}.
\newblock


\bibitem[Hafez et~al\mbox{.}(2014)]%
        {hafez2014bnem}
\bibfield{author}{\bibinfo{person}{Ahmed~Ibrahem Hafez}, \bibinfo{person}{Aboul~Ella Hassanien}, {and} \bibinfo{person}{Aly~A Fahmy}.} \bibinfo{year}{2014}\natexlab{}.
\newblock \showarticletitle{BNEM: a fast community detection algorithm using generative models}.
\newblock \bibinfo{journal}{\emph{Social Network Analysis and Mining}}  \bibinfo{volume}{4} (\bibinfo{year}{2014}), \bibinfo{pages}{1--20}.
\newblock


\bibitem[Haq and Wang(2016)]%
        {haq2016community}
\bibfield{author}{\bibinfo{person}{Nandinee Haq} {and} \bibinfo{person}{Z~Jane Wang}.} \bibinfo{year}{2016}\natexlab{}.
\newblock \showarticletitle{Community detection from genomic datasets across human cancers}. In \bibinfo{booktitle}{\emph{2016 IEEE Global Conference on Signal and Information Processing (GlobalSIP)}}. IEEE, \bibinfo{pages}{1147--1150}.
\newblock


\bibitem[He and Evans(2010)]%
        {he2010graph}
\bibfield{author}{\bibinfo{person}{Yong He} {and} \bibinfo{person}{Alan Evans}.} \bibinfo{year}{2010}\natexlab{}.
\newblock \showarticletitle{Graph theoretical modeling of brain connectivity}.
\newblock \bibinfo{journal}{\emph{Current opinion in neurology}} \bibinfo{volume}{23}, \bibinfo{number}{4} (\bibinfo{year}{2010}), \bibinfo{pages}{341--350}.
\newblock


\bibitem[Hesamipour et~al\mbox{.}(2022)]%
        {hesamipour2022detecting}
\bibfield{author}{\bibinfo{person}{Sajjad Hesamipour}, \bibinfo{person}{Mohammad~Ali Balafar}, \bibinfo{person}{Saeed Mousazadeh}, {et~al\mbox{.}}} \bibinfo{year}{2022}\natexlab{}.
\newblock \showarticletitle{Detecting communities in complex networks using an adaptive genetic algorithm and node similarity-based encoding}.
\newblock \bibinfo{journal}{\emph{Complexity}}  \bibinfo{volume}{2023} (\bibinfo{year}{2022}).
\newblock


\bibitem[Kim et~al\mbox{.}(2009)]%
        {kim2009centralized}
\bibfield{author}{\bibinfo{person}{Pan-Jun Kim}, \bibinfo{person}{Dong-Yup Lee}, {and} \bibinfo{person}{Hawoong Jeong}.} \bibinfo{year}{2009}\natexlab{}.
\newblock \showarticletitle{Centralized modularity of N-linked glycosylation pathways in mammalian cells}.
\newblock \bibinfo{journal}{\emph{PloS one}} \bibinfo{volume}{4}, \bibinfo{number}{10} (\bibinfo{year}{2009}), \bibinfo{pages}{e7317}.
\newblock


\bibitem[Kloster and Gleich(2014)]%
        {com-kloster14}
\bibfield{author}{\bibinfo{person}{K. Kloster} {and} \bibinfo{person}{D. Gleich}.} \bibinfo{year}{2014}\natexlab{}.
\newblock \showarticletitle{{Heat kernel based community detection}}. In \bibinfo{booktitle}{\emph{Proceedings of the 20th ACM SIGKDD international conference on Knowledge discovery and data mining}}. \bibinfo{publisher}{ACM}, \bibinfo{address}{New York, USA}, \bibinfo{pages}{1386--1395}.
\newblock


\bibitem[Kolodziej et~al\mbox{.}(2019)]%
        {suite19}
\bibfield{author}{\bibinfo{person}{S. Kolodziej}, \bibinfo{person}{M. Aznaveh}, \bibinfo{person}{M. Bullock}, \bibinfo{person}{J. David}, \bibinfo{person}{T. Davis}, \bibinfo{person}{M. Henderson}, \bibinfo{person}{Y. Hu}, {and} \bibinfo{person}{R. Sandstrom}.} \bibinfo{year}{2019}\natexlab{}.
\newblock \showarticletitle{{The SuiteSparse matrix collection website interface}}.
\newblock \bibinfo{journal}{\emph{The Journal of Open Source Software}} \bibinfo{volume}{4}, \bibinfo{number}{35} (\bibinfo{date}{Mar} \bibinfo{year}{2019}), \bibinfo{pages}{1244}.
\newblock


\bibitem[Kumar et~al\mbox{.}(2021)]%
        {kumar2021elpr}
\bibfield{author}{\bibinfo{person}{Sanjay Kumar}, \bibinfo{person}{Lakshay Singhla}, \bibinfo{person}{Kshitij Jindal}, \bibinfo{person}{Khyati Grover}, {and} \bibinfo{person}{BS Panda}.} \bibinfo{year}{2021}\natexlab{}.
\newblock \showarticletitle{IM-ELPR: Influence maximization in social networks using label propagation based community structure}.
\newblock \bibinfo{journal}{\emph{Applied Intelligence}} (\bibinfo{year}{2021}), \bibinfo{pages}{1--19}.
\newblock


\bibitem[Kuzmin et~al\mbox{.}(2015)]%
        {kuzmin2015parallelizing}
\bibfield{author}{\bibinfo{person}{Konstantin Kuzmin}, \bibinfo{person}{Mingming Chen}, {and} \bibinfo{person}{Boleslaw~K Szymanski}.} \bibinfo{year}{2015}\natexlab{}.
\newblock \showarticletitle{Parallelizing SLPA for scalable overlapping community detection}.
\newblock \bibinfo{journal}{\emph{Scientific Programming}}  \bibinfo{volume}{2015} (\bibinfo{year}{2015}), \bibinfo{pages}{4--4}.
\newblock


\bibitem[Leskovec(2021)]%
        {com-leskovec21}
\bibfield{author}{\bibinfo{person}{J. Leskovec}.} \bibinfo{year}{2021}\natexlab{}.
\newblock \bibinfo{title}{{CS224W: Machine Learning with Graphs | 2021 | Lecture 13.3 - Louvain Algorithm}}.
\newblock
\newblock
\urldef\tempurl%
\url{https://www.youtube.com/watch?v=0zuiLBOIcsw}
\showURL{%
\tempurl}


\bibitem[Leung et~al\mbox{.}(2009)]%
        {leung2009towards}
\bibfield{author}{\bibinfo{person}{Ian~XY Leung}, \bibinfo{person}{Pan Hui}, \bibinfo{person}{Pietro Lio}, {and} \bibinfo{person}{Jon Crowcroft}.} \bibinfo{year}{2009}\natexlab{}.
\newblock \showarticletitle{Towards real-time community detection in large networks}.
\newblock \bibinfo{journal}{\emph{Physical Review E}} \bibinfo{volume}{79}, \bibinfo{number}{6} (\bibinfo{year}{2009}), \bibinfo{pages}{066107}.
\newblock


\bibitem[Liu et~al\mbox{.}(2020)]%
        {com-liu20}
\bibfield{author}{\bibinfo{person}{X. Liu}, \bibinfo{person}{M. Halappanavar}, \bibinfo{person}{K. Barker}, \bibinfo{person}{A. Lumsdaine}, {and} \bibinfo{person}{A. Gebremedhin}.} \bibinfo{year}{2020}\natexlab{}.
\newblock \showarticletitle{{Direction-optimizing label propagation and its application to community detection}}. In \bibinfo{booktitle}{\emph{Proceedings of the 17th ACM International Conference on Computing Frontiers}}. \bibinfo{publisher}{ACM}, \bibinfo{address}{New York, NY, USA}, \bibinfo{pages}{192--201}.
\newblock
\showISBNx{9781450379564}


\bibitem[Luecken(2016)]%
        {luecken2016application}
\bibfield{author}{\bibinfo{person}{Malte Luecken}.} \bibinfo{year}{2016}\natexlab{}.
\newblock \emph{\bibinfo{title}{Application of multi-resolution partitioning of interaction networks to the study of complex disease}}.
\newblock \bibinfo{thesistype}{Ph.\,D. Dissertation}. \bibinfo{school}{University of Oxford}.
\newblock


\bibitem[Ma et~al\mbox{.}(2019)]%
        {ma2019comparative}
\bibfield{author}{\bibinfo{person}{Jun Ma}, \bibinfo{person}{Jenny Wang}, \bibinfo{person}{Laleh~Soltan Ghoraie}, \bibinfo{person}{Xin Men}, \bibinfo{person}{Benjamin Haibe-Kains}, {and} \bibinfo{person}{Penggao Dai}.} \bibinfo{year}{2019}\natexlab{}.
\newblock \showarticletitle{A comparative study of cluster detection algorithms in protein--protein interaction for drug target discovery and drug repurposing}.
\newblock \bibinfo{journal}{\emph{Frontiers in pharmacology}}  \bibinfo{volume}{10} (\bibinfo{year}{2019}), \bibinfo{pages}{109}.
\newblock


\bibitem[Newman(2006a)]%
        {com-newman06}
\bibfield{author}{\bibinfo{person}{M. Newman}.} \bibinfo{year}{2006}\natexlab{a}.
\newblock \showarticletitle{{Finding community structure in networks using the eigenvectors of matrices}}.
\newblock \bibinfo{journal}{\emph{Physical review E}} \bibinfo{volume}{74}, \bibinfo{number}{3} (\bibinfo{year}{2006}), \bibinfo{pages}{036104}.
\newblock


\bibitem[Newman(2006b)]%
        {newman2006modularity}
\bibfield{author}{\bibinfo{person}{Mark~EJ Newman}.} \bibinfo{year}{2006}\natexlab{b}.
\newblock \showarticletitle{Modularity and community structure in networks}.
\newblock \bibinfo{journal}{\emph{Proceedings of the national academy of sciences}} \bibinfo{volume}{103}, \bibinfo{number}{23} (\bibinfo{year}{2006}), \bibinfo{pages}{8577--8582}.
\newblock


\bibitem[Peng et~al\mbox{.}(2014)]%
        {peng2014accelerating}
\bibfield{author}{\bibinfo{person}{Chengbin Peng}, \bibinfo{person}{Tamara~G Kolda}, {and} \bibinfo{person}{Ali Pinar}.} \bibinfo{year}{2014}\natexlab{}.
\newblock \showarticletitle{Accelerating community detection by using k-core subgraphs}.
\newblock \bibinfo{journal}{\emph{arXiv preprint arXiv:1403.2226}} (\bibinfo{year}{2014}).
\newblock


\bibitem[Popa et~al\mbox{.}(2011)]%
        {popa2011directed}
\bibfield{author}{\bibinfo{person}{Ovidiu Popa}, \bibinfo{person}{Einat Hazkani-Covo}, \bibinfo{person}{Giddy Landan}, \bibinfo{person}{William Martin}, {and} \bibinfo{person}{Tal Dagan}.} \bibinfo{year}{2011}\natexlab{}.
\newblock \showarticletitle{Directed networks reveal genomic barriers and DNA repair bypasses to lateral gene transfer among prokaryotes}.
\newblock \bibinfo{journal}{\emph{Genome research}} \bibinfo{volume}{21}, \bibinfo{number}{4} (\bibinfo{year}{2011}), \bibinfo{pages}{599--609}.
\newblock


\bibitem[Raghavan et~al\mbox{.}(2007)]%
        {com-raghavan07}
\bibfield{author}{\bibinfo{person}{U. Raghavan}, \bibinfo{person}{R. Albert}, {and} \bibinfo{person}{S. Kumara}.} \bibinfo{year}{2007}\natexlab{}.
\newblock \showarticletitle{{Near linear time algorithm to detect community structures in large-scale networks}}.
\newblock \bibinfo{journal}{\emph{Physical Review E}} \bibinfo{volume}{76}, \bibinfo{number}{3} (\bibinfo{date}{Sep} \bibinfo{year}{2007}), \bibinfo{pages}{036106--1--036106--11}.
\newblock


\bibitem[Reichardt and Bornholdt(2006)]%
        {reichardt2006statistical}
\bibfield{author}{\bibinfo{person}{J{\"o}rg Reichardt} {and} \bibinfo{person}{Stefan Bornholdt}.} \bibinfo{year}{2006}\natexlab{}.
\newblock \showarticletitle{Statistical mechanics of community detection}.
\newblock \bibinfo{journal}{\emph{Physical review E}} \bibinfo{volume}{74}, \bibinfo{number}{1} (\bibinfo{year}{2006}), \bibinfo{pages}{016110}.
\newblock


\bibitem[Rivera et~al\mbox{.}(2010)]%
        {rivera2010nemo}
\bibfield{author}{\bibinfo{person}{Corban~G Rivera}, \bibinfo{person}{Rachit Vakil}, {and} \bibinfo{person}{Joel~S Bader}.} \bibinfo{year}{2010}\natexlab{}.
\newblock \showarticletitle{NeMo: network module identification in Cytoscape}.
\newblock \bibinfo{journal}{\emph{BMC bioinformatics}}  \bibinfo{volume}{11} (\bibinfo{year}{2010}), \bibinfo{pages}{1--9}.
\newblock


\bibitem[Rosvall and Bergstrom(2008)]%
        {com-rosvall08}
\bibfield{author}{\bibinfo{person}{M. Rosvall} {and} \bibinfo{person}{C. Bergstrom}.} \bibinfo{year}{2008}\natexlab{}.
\newblock \showarticletitle{{Maps of random walks on complex networks reveal community structure}}.
\newblock \bibinfo{journal}{\emph{Proceedings of the national academy of sciences}} \bibinfo{volume}{105}, \bibinfo{number}{4} (\bibinfo{year}{2008}), \bibinfo{pages}{1118--1123}.
\newblock


\bibitem[Sahu(2023a)]%
        {sahu2023gveleiden}
\bibfield{author}{\bibinfo{person}{Subhajit Sahu}.} \bibinfo{year}{2023}\natexlab{a}.
\newblock \showarticletitle{GVE-Leiden: Fast Leiden Algorithm for Community Detection in Shared Memory Setting}.
\newblock \bibinfo{journal}{\emph{arXiv preprint arXiv:2312.13936}} (\bibinfo{year}{2023}).
\newblock


\bibitem[Sahu(2023b)]%
        {sahu2023gvelouvain}
\bibfield{author}{\bibinfo{person}{S. Sahu}.} \bibinfo{year}{2023}\natexlab{b}.
\newblock \showarticletitle{GVE-Louvain: Fast Louvain Algorithm for Community Detection in Shared Memory Setting}.
\newblock \bibinfo{journal}{\emph{arXiv preprint arXiv:2312.04876}} (\bibinfo{year}{2023}).
\newblock


\bibitem[Sahu(2023c)]%
        {sahu2023gvelpa}
\bibfield{author}{\bibinfo{person}{Subhajit Sahu}.} \bibinfo{year}{2023}\natexlab{c}.
\newblock \showarticletitle{GVE-LPA: Fast Label Propagation Algorithm (LPA) for Community Detection in Shared Memory Setting}.
\newblock \bibinfo{journal}{\emph{arXiv preprint arXiv:2312.08140}} (\bibinfo{year}{2023}).
\newblock


\bibitem[Sahu(2023d)]%
        {sahu2023selecting}
\bibfield{author}{\bibinfo{person}{S. Sahu}.} \bibinfo{year}{2023}\natexlab{d}.
\newblock \showarticletitle{Selecting a suitable Parallel Label-propagation based algorithm for Disjoint Community Detection}.
\newblock \bibinfo{journal}{\emph{arXiv preprint arXiv:2301.09125}} (\bibinfo{year}{2023}).
\newblock


\bibitem[Sahu(2024)]%
        {sahu2024addressing}
\bibfield{author}{\bibinfo{person}{Subhajit Sahu}.} \bibinfo{year}{2024}\natexlab{}.
\newblock \showarticletitle{An Approach for Addressing Internally-Disconnected Communities in Louvain Algorithm}.
\newblock \bibinfo{journal}{\emph{arXiv preprint arXiv:2402.11454}} (\bibinfo{year}{2024}).
\newblock


\bibitem[Salath{\'e} and Jones(2010)]%
        {salathe2010dynamics}
\bibfield{author}{\bibinfo{person}{Marcel Salath{\'e}} {and} \bibinfo{person}{James~H Jones}.} \bibinfo{year}{2010}\natexlab{}.
\newblock \showarticletitle{Dynamics and control of diseases in networks with community structure}.
\newblock \bibinfo{journal}{\emph{PLoS computational biology}} \bibinfo{volume}{6}, \bibinfo{number}{4} (\bibinfo{year}{2010}), \bibinfo{pages}{e1000736}.
\newblock


\bibitem[Sattari and Zamanifar(2018)]%
        {com-sattari18}
\bibfield{author}{\bibinfo{person}{M. Sattari} {and} \bibinfo{person}{K. Zamanifar}.} \bibinfo{year}{2018}\natexlab{}.
\newblock \showarticletitle{{A spreading activation-based label propagation algorithm for overlapping community detection in dynamic social networks}}.
\newblock \bibinfo{journal}{\emph{Data \& knowledge engineering}}  \bibinfo{volume}{113} (\bibinfo{date}{Jan} \bibinfo{year}{2018}), \bibinfo{pages}{155--170}.
\newblock
\showISSN{0169023X}


\bibitem[Soman and Narang(2011)]%
        {soman2011fast}
\bibfield{author}{\bibinfo{person}{Jyothish Soman} {and} \bibinfo{person}{Ankur Narang}.} \bibinfo{year}{2011}\natexlab{}.
\newblock \showarticletitle{Fast community detection algorithm with gpus and multicore architectures}. In \bibinfo{booktitle}{\emph{2011 IEEE International Parallel \& Distributed Processing Symposium}}. IEEE, \bibinfo{pages}{568--579}.
\newblock


\bibitem[Souravlas et~al\mbox{.}(2021)]%
        {com-souravlas21}
\bibfield{author}{\bibinfo{person}{S. Souravlas}, \bibinfo{person}{A. Sifaleras}, \bibinfo{person}{M. Tsintogianni}, {and} \bibinfo{person}{S. Katsavounis}.} \bibinfo{year}{2021}\natexlab{}.
\newblock \showarticletitle{A classification of community detection methods in social networks: a survey}.
\newblock \bibinfo{journal}{\emph{International journal of general systems}} \bibinfo{volume}{50}, \bibinfo{number}{1} (\bibinfo{date}{Jan} \bibinfo{year}{2021}), \bibinfo{pages}{63--91}.
\newblock


\bibitem[Staudt et~al\mbox{.}(2016)]%
        {staudt2016networkit}
\bibfield{author}{\bibinfo{person}{C.L. Staudt}, \bibinfo{person}{A. Sazonovs}, {and} \bibinfo{person}{H. Meyerhenke}.} \bibinfo{year}{2016}\natexlab{}.
\newblock \showarticletitle{NetworKit: A tool suite for large-scale complex network analysis}.
\newblock \bibinfo{journal}{\emph{Network Science}} \bibinfo{volume}{4}, \bibinfo{number}{4} (\bibinfo{year}{2016}), \bibinfo{pages}{508--530}.
\newblock


\bibitem[Staudt and Meyerhenke(2015)]%
        {staudt2015engineering}
\bibfield{author}{\bibinfo{person}{Christian~L Staudt} {and} \bibinfo{person}{Henning Meyerhenke}.} \bibinfo{year}{2015}\natexlab{}.
\newblock \showarticletitle{Engineering parallel algorithms for community detection in massive networks}.
\newblock \bibinfo{journal}{\emph{IEEE Transactions on Parallel and Distributed Systems}} \bibinfo{volume}{27}, \bibinfo{number}{1} (\bibinfo{year}{2015}), \bibinfo{pages}{171--184}.
\newblock


\bibitem[Traag and {\v{S}}ubelj(2023)]%
        {traag2023large}
\bibfield{author}{\bibinfo{person}{V.A. Traag} {and} \bibinfo{person}{L. {\v{S}}ubelj}.} \bibinfo{year}{2023}\natexlab{}.
\newblock \showarticletitle{Large network community detection by fast label propagation}.
\newblock \bibinfo{journal}{\emph{Scientific Reports}} \bibinfo{volume}{13}, \bibinfo{number}{1} (\bibinfo{year}{2023}), \bibinfo{pages}{2701}.
\newblock


\bibitem[Traag et~al\mbox{.}(2019)]%
        {com-traag19}
\bibfield{author}{\bibinfo{person}{V. Traag}, \bibinfo{person}{L. Waltman}, {and} \bibinfo{person}{N. Eck}.} \bibinfo{year}{2019}\natexlab{}.
\newblock \showarticletitle{{From Louvain to Leiden: guaranteeing well-connected communities.}}
\newblock \bibinfo{journal}{\emph{Scientific Reports}} \bibinfo{volume}{9}, \bibinfo{number}{1} (\bibinfo{date}{Mar} \bibinfo{year}{2019}), \bibinfo{pages}{5233}.
\newblock


\bibitem[Whang et~al\mbox{.}(2013)]%
        {com-whang13}
\bibfield{author}{\bibinfo{person}{J. Whang}, \bibinfo{person}{D. Gleich}, {and} \bibinfo{person}{I. Dhillon}.} \bibinfo{year}{2013}\natexlab{}.
\newblock \showarticletitle{{Overlapping community detection using seed set expansion}}. In \bibinfo{booktitle}{\emph{Proceedings of the 22nd ACM international conference on Information \& Knowledge Management}}. \bibinfo{pages}{2099--2108}.
\newblock


\bibitem[Wolf et~al\mbox{.}(2019)]%
        {wolf2019paga}
\bibfield{author}{\bibinfo{person}{F~Alexander Wolf}, \bibinfo{person}{Fiona~K Hamey}, \bibinfo{person}{Mireya Plass}, \bibinfo{person}{Jordi Solana}, \bibinfo{person}{Joakim~S Dahlin}, \bibinfo{person}{Berthold G{\"o}ttgens}, \bibinfo{person}{Nikolaus Rajewsky}, \bibinfo{person}{Lukas Simon}, {and} \bibinfo{person}{Fabian~J Theis}.} \bibinfo{year}{2019}\natexlab{}.
\newblock \showarticletitle{PAGA: graph abstraction reconciles clustering with trajectory inference through a topology preserving map of single cells}.
\newblock \bibinfo{journal}{\emph{Genome biology}}  \bibinfo{volume}{20} (\bibinfo{year}{2019}), \bibinfo{pages}{1--9}.
\newblock


\bibitem[Xie et~al\mbox{.}(2013)]%
        {com-xie13}
\bibfield{author}{\bibinfo{person}{J. Xie}, \bibinfo{person}{M. Chen}, {and} \bibinfo{person}{B. Szymanski}.} \bibinfo{year}{2013}\natexlab{}.
\newblock \showarticletitle{{LabelrankT: Incremental community detection in dynamic networks via label propagation}}. In \bibinfo{booktitle}{\emph{Proceedings of the Workshop on Dynamic Networks Management and Mining}}. \bibinfo{publisher}{ACM}, \bibinfo{address}{New York, USA}, \bibinfo{pages}{25--32}.
\newblock


\bibitem[Xie et~al\mbox{.}(2011)]%
        {com-xie11}
\bibfield{author}{\bibinfo{person}{J. Xie}, \bibinfo{person}{B. Szymanski}, {and} \bibinfo{person}{X. Liu}.} \bibinfo{year}{2011}\natexlab{}.
\newblock \showarticletitle{{SLPA: Uncovering overlapping communities in social networks via a speaker-listener interaction dynamic process}}. In \bibinfo{booktitle}{\emph{IEEE 11th International Conference on Data Mining Workshops}}. IEEE, \bibinfo{publisher}{IEEE}, \bibinfo{address}{Vancouver, Canada}, \bibinfo{pages}{344--349}.
\newblock


\bibitem[Xie and Szymanski(2011)]%
        {xie2011community}
\bibfield{author}{\bibinfo{person}{Jierui Xie} {and} \bibinfo{person}{Boleslaw~K Szymanski}.} \bibinfo{year}{2011}\natexlab{}.
\newblock \showarticletitle{Community detection using a neighborhood strength driven label propagation algorithm}. In \bibinfo{booktitle}{\emph{2011 IEEE Network Science Workshop}}. IEEE, \bibinfo{pages}{188--195}.
\newblock


\bibitem[Xing et~al\mbox{.}(2014)]%
        {com-xing14}
\bibfield{author}{\bibinfo{person}{Y. Xing}, \bibinfo{person}{F. Meng}, \bibinfo{person}{Y. Zhou}, \bibinfo{person}{M. Zhu}, \bibinfo{person}{M. Shi}, {and} \bibinfo{person}{G. Sun}.} \bibinfo{year}{2014}\natexlab{}.
\newblock \showarticletitle{{A node influence based label propagation algorithm for community detection in networks}}.
\newblock \bibinfo{journal}{\emph{The Scientific World Journal}}  \bibinfo{volume}{2014} (\bibinfo{year}{2014}), \bibinfo{pages}{1--14}.
\newblock


\bibitem[You et~al\mbox{.}(2020)]%
        {com-you20}
\bibfield{author}{\bibinfo{person}{X. You}, \bibinfo{person}{Y. Ma}, {and} \bibinfo{person}{Z. Liu}.} \bibinfo{year}{2020}\natexlab{}.
\newblock \showarticletitle{{A three-stage algorithm on community detection in social networks}}.
\newblock \bibinfo{journal}{\emph{Knowledge-Based Systems}}  \bibinfo{volume}{187} (\bibinfo{year}{2020}), \bibinfo{pages}{104822}.
\newblock


\bibitem[Zarayeneh and Kalyanaraman(2021)]%
        {com-zarayeneh21}
\bibfield{author}{\bibinfo{person}{N. Zarayeneh} {and} \bibinfo{person}{A. Kalyanaraman}.} \bibinfo{year}{2021}\natexlab{}.
\newblock \showarticletitle{{Delta-Screening: A Fast and Efficient Technique to Update Communities in Dynamic Graphs}}.
\newblock \bibinfo{journal}{\emph{IEEE transactions on network science and engineering}} \bibinfo{volume}{8}, \bibinfo{number}{2} (\bibinfo{date}{Apr} \bibinfo{year}{2021}), \bibinfo{pages}{1614--1629}.
\newblock


\end{thebibliography}

\clearpage
\appendix
\section{Appendix}
\subsection{Finding disconnected communities}

We present a parallel algorithm aimed at identifying disconnected communities using the original graph and vertex community memberships as input. The fundamental principle involves evaluating each community's size, selecting a representative vertex, traversing within the community while avoiding neighboring communities, and flagging a community as disconnected if all its vertices are unreachable. We explore four distinct approaches, characterized by their utilization of parallel Depth-First Search (DFS) or Breadth-First Search (BFS), and whether per-thread or shared \textit{visited} flags are employed. In the case of shared visited flags, each thread scans all vertices but exclusively processes its assigned community based on the community ID. Our results demonstrate that employing parallel BFS traversal with a shared flag vector yields the most efficient outcomes. Due to the deterministic nature of this algorithm, all approaches produce identical results. Algorithm \ref{alg:disconnected} presents the pseudocode for this approach, where the \texttt{disconnectedCommunities()} function takes the input graph $G$ and community membership $C$, returning the disconnected flag $D$ for each community.

\begin{algorithm}[hbtp]
\caption{Finding disconnected communities in parallel \cite{sahu2023gveleiden}.}
\label{alg:disconnected}
\begin{algorithmic}[1]
\Require{$G(V, E)$: Input graph}
\Require{$C$: Community membership of each vertex}
\Ensure{$D$: Disconnected flag for each community}
\Ensure{$S$: Size of each community}
\Ensure{$f_{if}$: Perform BFS to vertex $j$ if condition satisfied}
\Ensure{$f_{do}$: Perform operation after each vertex is visited}
\Ensure{$reached$: Number of vertices reachable from $i$ to $i$'s community}
\Ensure{$vis$: Visited flag for each vertex}
\Ensure{$work_t$: Work-list of current thread}

\Statex

\Function{disconnectedCommunities}{$G, C$} \label{alg:disconnected--begin}
  \State $D \gets \{\}$ \textbf{;} $vis \gets \{\}$ \label{alg:disconnected--init}
  \State $S \gets communitySizes(G, C)$ \label{alg:disconnected--sizes}
  \ForAll{\textbf{threads in parallel}} \label{alg:disconnected--threads-begin}
    \ForAll{$i \in V$} \label{alg:disconnected--loop-begin}
      \State $c \gets C[i]$ \textbf{;} $reached \gets 0$ \label{alg:disconnected--unreached}
      \State $\rhd$ Skip if community $c$ is empty, or
      \State $\rhd$ does not belong to work-list of current thread.
      \If{$S[c] = 0$ \textbf{or} $c \notin work_t$} \textbf{continue} \label{alg:disconnected--work}
      \EndIf
      \State $f_{if} \gets (j) \implies C[j] = c$
      \State $f_{do} \gets (j) \implies reached \gets reached + 1$
      \State $bfsVisitForEach(vis, G, i, f_{if}, f_{do})$ \label{alg:disconnected--bfs}
      \If{$reached < S[c]$} $D[c] \gets 1$ \label{alg:disconnected--mark}
      \EndIf
      \State $S[c] \gets 0$ \label{alg:disconnected--processed}
    \EndFor \label{alg:disconnected--loop-end}
  \EndFor \label{alg:disconnected--threads-end}
  \Return{$D$}
\EndFunction \label{alg:disconnected--end}
\end{algorithmic}
\end{algorithm}

Let us now explain Algorithm \ref{alg:disconnected}. Initially, in line \ref{alg:disconnected--init}, we set up the disconnected community flag $D$ and the visited vertices flags $vis$. Line \ref{alg:disconnected--sizes} calculates the size of each community $S$ concurrently using the \texttt{communitySizes()} function. Subsequently, each thread processes every vertex $i$ in the graph $G$ concurrently (lines \ref{alg:disconnected--loop-begin}-\ref{alg:disconnected--loop-end}). In line \ref{alg:disconnected--unreached}, we ascertain the community membership of $i$ ($c$) and initialize the count of reached vertices from $i$ to $0$. If community $c$ is either empty or not in the work-list of the current thread $work_t$, the thread proceeds to the next iteration (line \ref{alg:disconnected--work}). However, if community $c$ is non-empty and in the work-list of the current thread $work_t$, we execute BFS from vertex $i$ to explore vertices within the same community. This involves utilizing lambda functions $f_{if}$ to conditionally execute BFS to vertex $j$ if it belongs to the same community and $f_{do}$ to update the count of reached vertices after each vertex is visited during BFS (line \ref{alg:disconnected--bfs}). If the number of vertices $reached$ during BFS is less than the community size $S[c]$, we flag community $c$ as disconnected (line \ref{alg:disconnected--mark}). Finally, we update the size of the community $S[c]$ to $0$, indicating that the community has been processed (line \ref{alg:disconnected--processed}). It's worth noting that the work-list $work_t$ for each thread with ID $t$ is defined as a set containing communities $[t\chi,\ t(\chi+1))\ \cup\ [T\chi + t\chi,\ T\chi + t(\chi+1))\ \cup\ \ldots$, where $\chi$ is the chunk size and $T$ is the number of threads. In our implementation, we use a chunk size of $\chi = 1024$.

\subsection{Additional Performance comparison}
\label{sec:comparison-extra}

\begin{figure*}[hbtp]
  \centering
  \subfigure[Runtime in seconds (logarithmic scale) with \textit{GVE-LPA} and \textit{GSL-LPA}]{
    \label{fig:cmpduo--runtime}
    \includegraphics[width=0.98\linewidth]{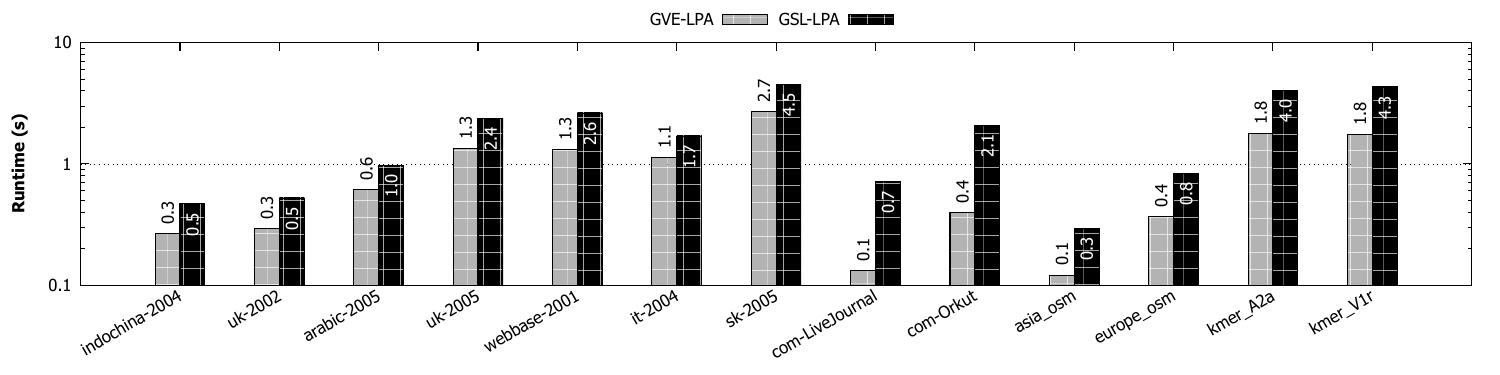}
  } \\[-0ex]
  \subfigure[Speedup of \textit{GVE-LPA} with respect to \textit{GSL-LPA}.]{
    \label{fig:cmpduo--speedup}
    \includegraphics[width=0.98\linewidth]{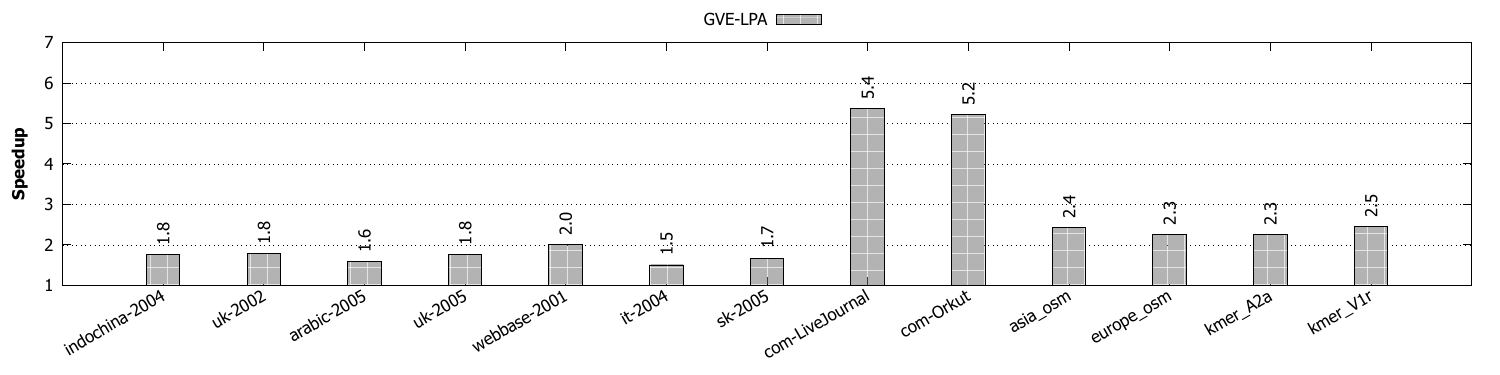}
  } \\[-0ex]
  \subfigure[Modularity of communities obtained with \textit{GVE-LPA} and \textit{GSL-LPA}.]{
    \label{fig:cmpduo--modularity}
    \includegraphics[width=0.98\linewidth]{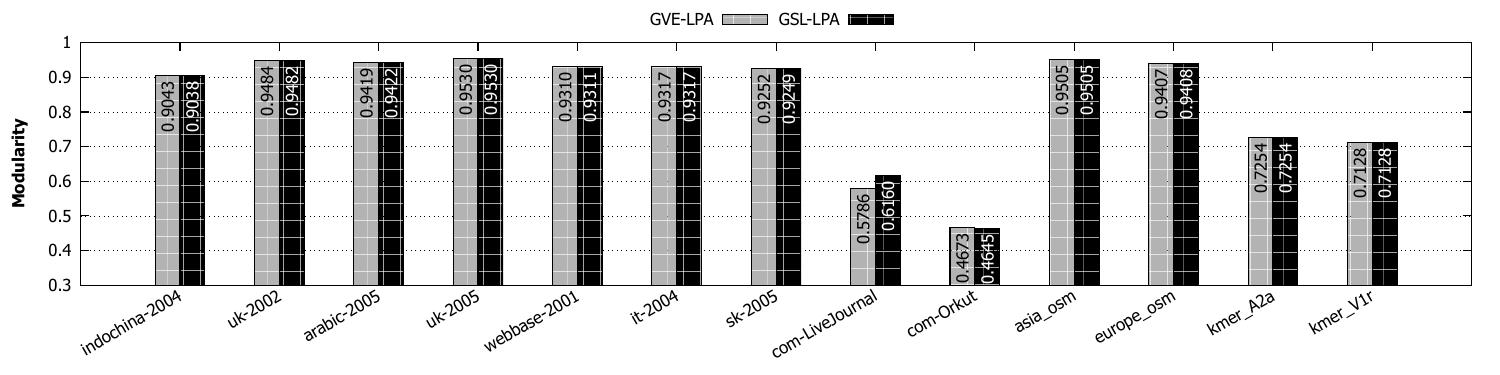}
  } \\[-0ex]
  \subfigure[Fraction of disconnected communities (logarithmic scale) with \textit{GVE-LPA} and \textit{GSL-LPA}.]{
    \label{fig:cmpduo--disconnected}
    \includegraphics[width=0.98\linewidth]{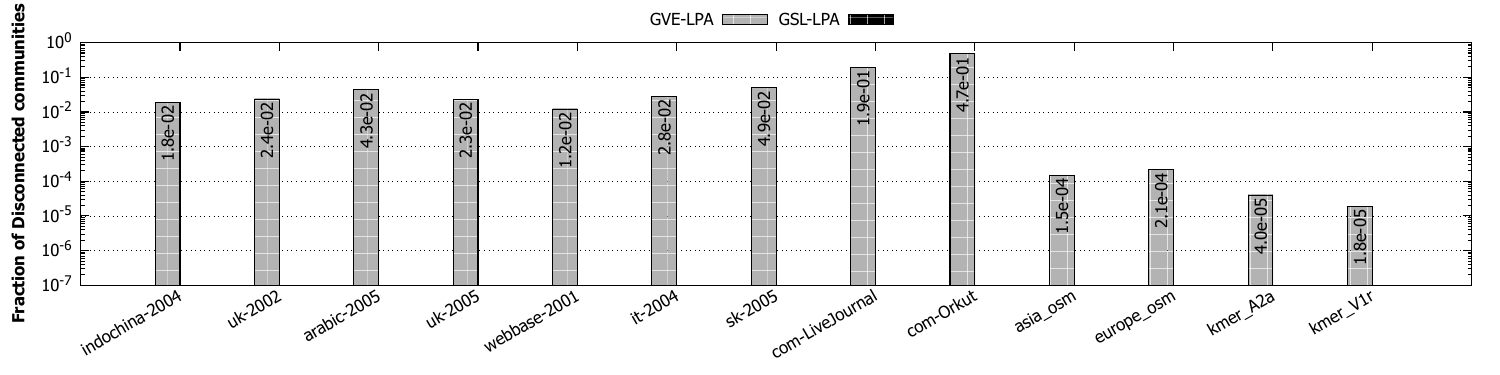}
  } \\[-2ex]
  \caption{Runtime in seconds (log-scale), speedup, modularity, and fraction of disconnected communities (log-scale) with \textit{GVE-LPA} \cite{sahu2023gvelpa} and \textit{GSL-LPA} for each graph in the dataset.}
  \label{fig:cmpduo}
\end{figure*}

Next, we proceed with the comparison of performance between GSL-LPA and GVE-LPA \cite{sahu2023gvelpa}. Following a similar methodology as earlier, we run both implementations five times for every graph in the dataset to reduce measurement noise, and present the averages in Figures \ref{fig:cmpduo--runtime}, \ref{fig:cmpduo--speedup}, \ref{fig:cmpduo--modularity}, and \ref{fig:cmpduo--disconnected}.

Figure \ref{fig:cmpduo--runtime} illustrates the runtimes of GVE-LPA and GSL-LPA on each graph in the dataset. On average, GSL-LPA demonstrates a runtime approximately $125\%$ longer than GVE-LPA. This additional computational time represents a trade-off made to ensure the absence of internally disconnected communities. To minimize the runtime for GSL-LPA, it is necessary to optimize Algorithm \ref{alg:splitbfs}. We plan to address this in the future.

Figure \ref{fig:cmpduo--modularity} showcases the modularity of communities obtained by each implementation. On average, GSL-LPA achieves a $0.4\%$ higher modularity compared to GVE-LPA. Finally, Figure \ref{fig:cmpduo--disconnected} presents the fraction of internally disconnected communities identified by each implementation. Communities obtained with GSL-LPA exhibit no disconnected communities, while those identified with GVE-LPA feature an average of $6.6\%$ disconnected communities.

\ignore{Next, we proceed to compare the performance of GVE-LPA with GVE-Louvain in Figure \ref{fig:louvainrak-compare}. Figures \ref{fig:louvainrak-compare--runtime}, \ref{fig:louvainrak-compare--speedup}, and \ref{fig:louvainrak-compare--modularity} present the runtimes, speedup (of GVE-LPA with respect to GVE-Louvain), and modularity of GVE-Louvain and GVE-LPA for each graph in the dataset. GVE-LPA offers a mean speedup of $5.4\times$ compared to GVE-Louvain, particularly on social networks, road networks, and protein k-mer graphs, while achieving on average $10.9\%$ lower modularity, especially on social networks and protein k-mer graphs.}

\end{document}